\documentclass[aoas]{imsart}

\RequirePackage{amsthm,amsmath,amsfonts,amssymb,bm,mathtools}
\RequirePackage[authoryear]{natbib}
\RequirePackage[colorlinks,citecolor=blue,urlcolor=blue]{hyperref}
\RequirePackage{graphicx,multirow,booktabs}
\RequirePackage{utfsym}
\usepackage{ulem}
\newcommand{\dif}{\mathop{}\!\mathrm{d}}

\startlocaldefs
\theoremstyle{plain}

\theoremstyle{remark}

\endlocaldefs

\begin{document}

\begin{frontmatter}
\title{Measuring Information Transfer Between Nodes in a Brain Network through Spectral Transfer Entropy}
\runtitle{Spectral Transfer Entropy}

\begin{aug}
\author{\fnms{Paolo~Victor}~\snm{Redondo}\ead[label=e1]{paolovictor.redondo@kaust.edu.sa}},
\author{\fnms{Rapha\"el}~\snm{Huser}\ead[label=e2]{raphael.huser@kaust.edu.sa}\orcid{0000-0002-1228-2071}}
\and
\author{\fnms{Hernando}~\snm{Ombao}\ead[label=e3]{hernando.ombao@kaust.edu.sa}}

\address{Statistics Program, Computer, Electrical, and Mathematical Science and Engineering (CEMSE) Division, King Abdullah University of Science and Technology (KAUST), Thuwal 23955-6900, Saudi Arabia\printead[presep={,\\ }]{e1,e2,e3}}

\end{aug}

\begin{abstract}
Brain connectivity characterizes interactions between different regions of a brain network during resting-state or performance of a cognitive task. In studying brain signals such as electroencephalograms (EEG), one formal approach to investigating connectivity is through an information-theoretic causal measure called transfer entropy (TE). To enhance the functionality of TE in brain signal analysis, we propose a novel methodology that captures cross-channel information transfer in the frequency domain. Specifically, we introduce a new measure, the \textit{spectral transfer entropy (STE)}, to quantify the magnitude and direction of information flow from a band-specific oscillation of one channel to another band-specific oscillation of another channel. The main advantage of our proposed approach is that it formulates TE in a novel way to perform inference on band-specific oscillations while maintaining robustness to the inherent problems associated with filtering. In addition, an advantage of STE is that it allows adjustments for multiple comparisons to control false positive rates. Another novel contribution is a simple yet efficient method for estimating STE using vine copula theory. This method can produce an exact zero estimate of STE (which is the boundary point of the parameter space) without the need for bias adjustments. With the vine copula representation, a null copula model, which exhibits zero STE, is defined, thus enabling straightforward significance testing through standard resampling. Lastly, we demonstrate the advantage of the proposed STE measure through numerical experiments and provide interesting and novel findings on the analysis of EEG data in a visual-memory experiment.
\end{abstract}


\end{frontmatter}

\section{Introduction}\label{chap:introduction}

There are two general objectives when studying brain connectivity: establishing either (i) \emph{functional connectivity}, which characterizes the statistical dependence between different brain regions; or (ii) \emph{effective connectivity}, which characterizes the causal relationships reflecting the direction of information transfer across nodes in the entire brain network \citep{cao2022brain}. Although functional connectivity approaches such as the classical Fourier coherence and wavelet coherence analysis provide practical insights on various brain dynamics (see for example \cite{sakkalis2006time,sankari2012wavelet,park2014estimating,fiecas2016modeling,khan2022development}), one drawback of these methods is that the derived connectivity does not specify the \textit{direction} of information from one node to another. This limits the type of neurophysiological interpretations that can be drawn by practitioners. In \cite{reid2019advancing}, a framework was proposed where association from functional connectivity approaches may be extended to causation. However, the class of effective connectivity methods remains more popular as they naturally incorporate directionality in the derived brain networks.

\begin{figure}[hbt!]
	\centerline{
		\includegraphics[width=0.75\textwidth]{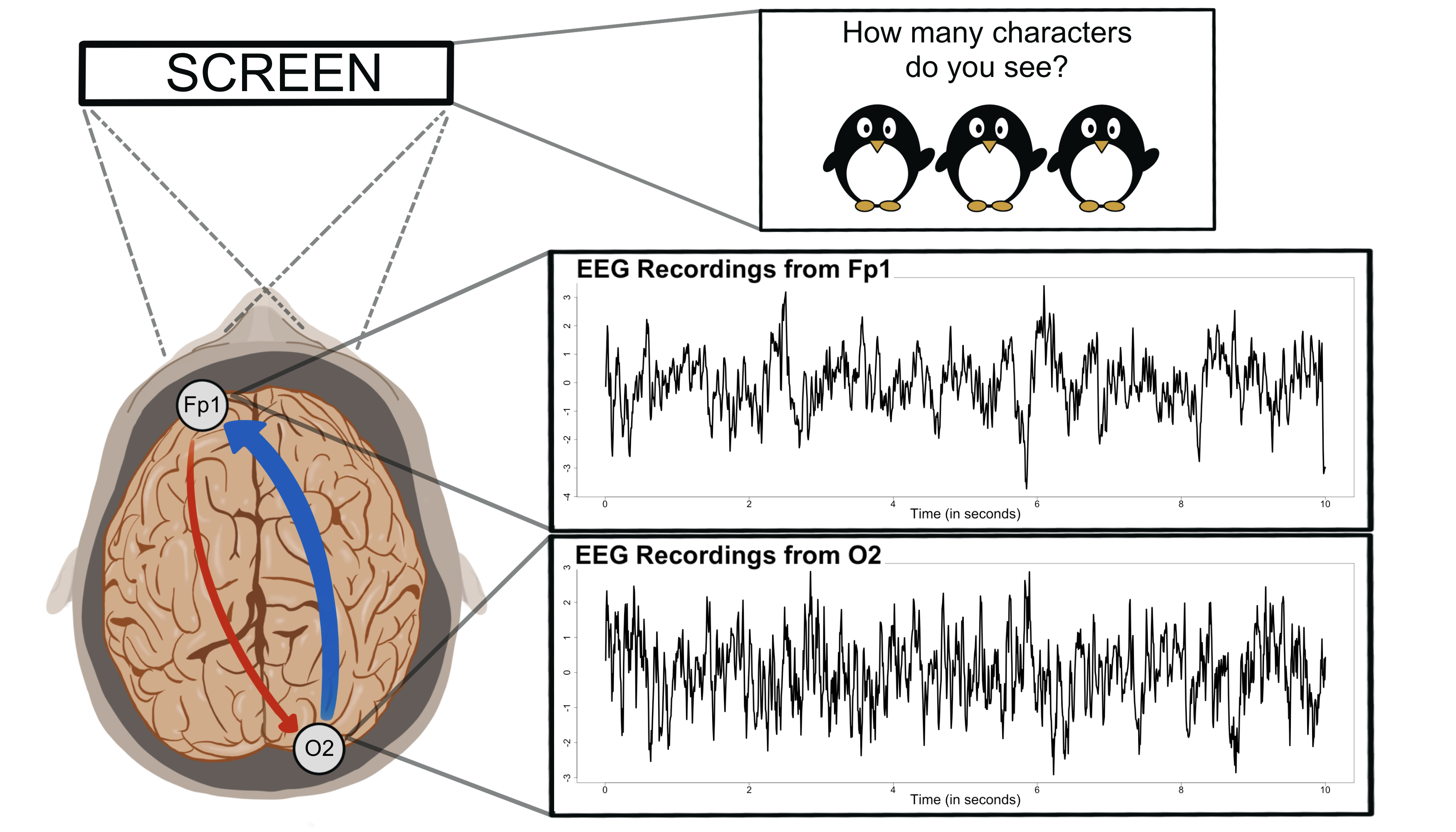}
	}
	\caption{EEG recordings at specific channels from a child participant during a visual-memory experiment; and the hypothesized causal relationship between two brain regions.}
	\label{fig:eeg_visual}
\end{figure}
From the cortical electrical activity recorded through electroencephalography (EEG), our primary goal in this paper is to derive the effective brain connectivity structure that characterizes information flow between nodes in the brain network during a cognitive experiment. More formally, consider EEG recordings from two channels $X$ and $Y$. Our objective is to measure the magnitude and test the significance of the causal impact of the extracted oscillations from $X$ to $Y$ (and vice versa); see Figure~\ref{fig:eeg_visual} to describe the motivation of our work. Another scientific interest is to compare effective brain networks between different groups (e.g., patient vs control). Thus, we aim to develop a statistical method that can determine differences in causal links (absence of links; increased or decreased magnitudes) between the healthy vs. disease populations to deepen our understanding of cognitive dysfunctions associated with the disorder. 

Granger causality (GC) is a widely-used causal inference framework to test hypotheses on brain networks which we briefly describe. Consider two time series $X$ and $Y$. Here, ``$X$ Granger-causes $Y$'' implies that the variance (or uncertainty) of the predictions for $Y$ is reduced when the predictor includes the past values of $X$ in addition to its own history \citep{granger1963economic,granger1969investigating}. For a review on the recent advancements on GC, see \cite{shojaie2022granger}. Since GC suggests a directional relationship, it offers an elegant foundation for exploring information transfer between signals coming from different brain regions. However, despite its well-built theoretical justifications, several commonly-made restrictive assumptions (e.g., linearity, additivity and Gaussianity) hinders the application of GC to brain functional data \citep{stokes2017study}.

Another framework for investigating effective brain connectivity is the dynamic causal modeling (DCM) \citep{friston2003dynamic} which is a model-based technique that tests different hypotheses of connectivity using Bayesian model comparison to find the most likely network that generated the observed signals \citep{van2013nonlinear}. This allows for characterizing the dynamics of causal interactions between different brain regions and has contributed to advancements of functional magnetic resonance imaging (fMRI) \citep{stephan2008nonlinear,havlicek2015physiologically,frassle2021regression} and EEG analysis \citep{kiebel2006dynamic,brown2012dynamic}. However, DCM requires concrete prior knowledge of the driver that causes changes in the network structure. Hence, it is most suitable for the task-related experiment paradigm \citep{saetia2021constructing}. This limits the utility of DCM in performing an exploratory analysis to continuously recorded brain signals, similar to the data we investigate in this paper.

Conceptualized by \cite{schreiber2000measuring}, an alternative approach that overcomes these limitations is through the concept of transfer entropy. Transfer entropy (TE) is an information-theoretic measure that quantifies the causal impact of $X$ to $Y$ by calculating the dependence between the current value $Y_t$ and the history of the other series, denoted by $\boldsymbol{X}_{t-k}$, given its own past $\boldsymbol{Y}_{t-\ell}$, for some positive lags $k$ and $\ell$, directly from their conditional distribution. Being equivalent to conditional mutual information (CMI), another dependence measure in information theory \citep{gray2011entropy,cover2012elements}, TE shares the same properties as CMI. For example, the interpretation of ``zero'' TE from $X$ to $Y$ implies conditional independence between $Y_t$ and the lagged values $\boldsymbol{X}_{t-k}$ given its own lagged values $\boldsymbol{Y}_{t-\ell}$. Moreover, TE does not assume any type of relationship (e.g., linear) between the series or impose any assumption on the distribution (e.g., Gaussian) of the series as the dependence it captures is modeled directly from joint and conditional distributions. Thus, one advantage of TE is that it offers a more flexible causal framework than GC and DCM. Consequently, it is more suitable in exploring effective brain connectivity \citep{vicente2011transfer}. For examples of TE application in brain data analysis, see \cite{huang2015identifying}, \cite{saetia2021constructing}, and \cite{sanjari2021assessment}.

In EEG analysis, a common objective is to conduct statistical inference on signals by decomposing (or filtering) into different frequency bands (e.g., delta (0.5–4 Hz), theta (4–8 Hz), alpha (8–12 Hz), beta (12–30 Hz), and gamma (30–100 Hz); see \cite{ombao2005slex,nunez2016electroencephalography,guerrero2023conex,ombao2022spectral}), which have well-known associated cognitive functions \citep{harmony2013functional,bjorge2017identification}. Recent spectral methods avoid the use of filtering (i.e., ``frequency \textit{band}-specific'' analysis) of potential distortion on temporal dependence and inability to isolate spectral causality \citep{florin2010effect,barnett2011behaviour,seth2015granger}. Many methods are based on the actual spectral representations of the series and perform a ``frequency-specific'' analysis at \textit{individual} frequencies. \cite{chen2019transfer} embedded the phase-space reconstruction method with the two-dimensional Fourier transform, while \cite{tian2021fourier} developed the theory for using wavelet transformation on the signals before calculating TE. However, these frequency-specific approaches often result in two possible drawbacks: (i) increased false positive rate in detecting significant spectral causal influence; and (ii) vague practical interpretations.

Given that frequency-specific methods estimate TE for a large number of possible frequencies and for multiple pairs of signal sources, detected significant quantities may be inappropriate even after adjusting for multiple comparisons. In particular, when the method employs computationally intensive techniques for testing significance (such as resampling), it becomes burdensome as the required number of replicates per test greatly increases with the number of simultaneous comparisons being tested (to ensure an acceptable precision). In addition, when significance is determined only at a specific frequency in the boundary of two frequency bands, linking such results to established findings may be awkward and subjective. Motivated by the minimized utility of these existing tools, we pursue the use of frequency \textit{band}-specific filtered signals in formulating TE in the frequency domain and develop a methodology that yields results which are easy to interpret, while appropriately controlling the false positive rate in the brain connectivity analysis.

One attempt to define information transfer between band-specific oscillations is given in \cite{dimitriadis2016revealing}, employing the so-called neural gas (NG) algorithm to convert the signals into sequences of symbols, which are consequently used to calculate the symbolic transfer entropy (symTE) measure \citep{staniek2008symbolic}. Since the NG algorithm relies on an artificial neural network to produce the symbolic representations for the signals \citep{martinetz1993neural}, one problem of this approach is the lack of interpretability of the captured causal influence. Because the symbols are complex summarized versions of the original oscillations, the information transfer measured by symTE may not only be reflective the interactions of the filtered oscillations but also includes artifacts induced by the transformation. Another limitation is that the analysis \cite{dimitriadis2016revealing} conducted involves narrow frequency bands, i.e., theta (5-6 Hz) and high alpha (10-13 Hz), which resembles regularly fluctuating sinusoids (with minimal phase distortion), and down-sampling of higher frequencies, and hence, may not perform well when considering a wider coverage of frequencies such as the widely-used canonical frequency bands.

In this paper, we thus formulate TE to conduct inference on general band-specific oscillations with the corresponding interpretations naturally bounded to the specified frequencies. Our novel framework quantifies the amount and direction of information transfer between two time series from a ``frequency band'' perspective. Instead of directly calculating TE between two filtered band-specific signals, we develop a new spectral causal measure, which we call the \textit{spectral transfer entropy (STE)}, based on the maximum magnitude of the zero-mean filtered series within non-overlapping time blocks. This approach offers a new perspective on capturing causal relationships in a brain network which is, more importantly, robust to the inherent problems of filtering. To minimize the impact of having different marginal characteristics in estimation, we exploit the relationship, proven by \cite{ma2011mutual}, between copulas and the information-theoretic measure. More specifically, we develop an estimation procedure via vine copula models after expressing TE in terms of copula densities. By strategically arranging the variables in a D-vine structure \citep{kurowicka2005distribution}, one clear advantage of our method is that it leads to a simpler re-expression for calculating TE which can capture the boundary value of zero. Moreover, since it is easy to generate new observations from an estimated vine copula model, we develop a convenient resampling method for measuring uncertainty of the estimates that can readily accommodate correction for multiple comparisons.

The remainder of this paper is organized as follows. Section~\ref{chap:background} provides a brief review of copula theory, dependence measures in information theory and their close link. In Section~\ref{chap:methodology}, we propose our new causal measure in the frequency domain, the STE, and develop a novel estimation method based on vine copula models. To provide evidence on the performance of STE in capturing spectral influence, in Section~\ref{chap:numexp}, we use the proposed metric on simulated series. In Section~\ref{chap:analysis}, we apply the proposed STE method to analyze actual EEG recordings from multiple children with or without attention deficit hyperactivity disorder (ADHD). The STE method produced interesting findings and novel results that describe the effective brain connectivity during performance of a visual task. Finally, concluding notes and future directions of our work are discussed in Section~\ref{chap:conclusion}.

\section{Quantifying Dependence via Copula Theory and Information Theory}\label{chap:background}

Suppose that $\bm{X}$ is a $d$-dimensional random vector where $F(\bm{x})$ and $f(\bm{x})$ represent its cumulative distibution function (CDF) and probability density function (PDF), respectively. Similar notations are used for other random vectors when the necessity occurs. Here, we present two concepts that characterize dependence between variables, namely, copula theory and information theory. We first describe both concepts based on general random vectors but establish their relationship in the context of time series data afterwards. 

\subsection{Copula Theory and Vine Copula Models}\label{subchap:copula}
For any $d$-dimensional CDF $F$ with continuous univariate margins $F_1,\ldots,F_d$, Sklar's Theorem \citep{sklar1959fonctions} states that there exists a unique $d$-dimensional copula $C$ (i.e., a joint distribution function with uniform Unif$(0,1)$ margins) such that
\begin{equation}
  F(\bm{x}) = C\{F_1(x_1),\ldots,F_d(x_d)\}, ~~~\bm{x} \in \mathbb{R}^d.
  \label{eq:cop1}
\end{equation}
\noindent An implication of Equation~(\ref{eq:cop1}) is that, if the CDF $F$ is absolutely continuous, then the joint PDF $f$ can be expressed as
\begin{equation}
  f(\bm{x}) = c\{F_1(x_1),\ldots,F_d(x_d)\} \prod^{d}_{i=1} f_i(x_i),
  \label{eq:cop2}
\end{equation}
\noindent where $c(\cdot)$ is the corresponding copula density. By Equation~(\ref{eq:cop2}), the dependence structure of any jointly-distributed continuous random variables can be extracted independently from the marginal distributions. Moreover, given the margins $F_i(x_i), i = 1, \ldots, d$, \cite{aas2009pair} showed that any $d$-dimensional copula density can be decomposed as product of bivariate copulas and conditonal bivariate copulas. Although the decomposition is not unique, a graphical method, called regular vines, offers a systematic approach for specifying a valid decomposition \citep{bedford2001probability}.

In this work, we focus on a specific vine structure called the D-vine \citep{kurowicka2005distribution}, because it leads to simplifications for calculating the causal measure of interest, as explained in Section~\ref{subchap:vines} below. Assuming a D-vine structure, any $d$-dimensional copula density, $c(F_1(x_1),\ldots,F_d(x_d))$, can be written through conditional bivariate copulas as 
{
\footnotesize
\begin{equation*}
     c(F_1(x_1),...,F_d(x_d)) = \prod^{d-1}_{j=1} \prod^{d-j}_{i=1} c_{i,i+j | i+1, ..., i+j-1}(F(x_i | x_{i+1},...,x_{i+j-1}),F(x_{i+j} | x_{i+1},...,x_{i+j-1})).
\end{equation*}
}
\noindent Under the widely-used so-called ``simplifying assumption'' \citep{haff2010simplified}, which assumes that the bivariate conditional copulas involved do not depend on the conditioning variables, we obtain a valid $d$-dimensional copula model, which remains very flexible as it spans a rich class of copulas and that is also both easier to specify and faster to compute. Such simplified D-vine copulas may be expressed as
\begin{equation}
    c(F_1(x_1),\ldots,F_d(x_d)) = \prod^{d-1}_{j=1} \prod^{d-j}_{i=1} c_{i,i+j}(F(x_i | x_{i+1},\ldots,x_{i+j-1}),F(x_{i+j} | x_{i+1},\ldots,x_{i+j-1})).
    \label{eq:vine}
\end{equation}
\noindent Moreover, by strategically organizing the arrangement of variables in a D-vine of the form~(\ref{eq:vine}), the copula density of any subset $\{X_{j_1},\ldots,X_{j_n}~\mid~j_1,\ldots,j_n \in \{1,\ldots,d\},n \leq d\}$ can be obtained from Equation~(\ref{eq:vine}) by extracting the corresponding bivariate copula terms (see, for example, \citealp{czado2022vine}, for more details on vine copula modeling). With this, computations for some information-theoretic measures are simplified, which we exploit in our proposed estimation procedure for the STE in Section~\ref{subchap:TEestimation}.

\subsection{Mutual Information and Transfer Entropy}\label{subchap:MIandTE}
Mutual information (MI) is the key formulation of dependence in information theory (see \citealp{cover2012elements}). More formally, the mutual information between two random vectors $\bm{X}$ and $\bm{Y}$, denoted by $I(\bm{X},\bm{Y})$ is defined as
\begin{equation}
    I(\bm{X},\bm{Y}) = \iint f(\bm{x},\bm{y})\log \frac{f(\bm{x},\bm{y})}{f(\bm{x})f(\bm{y})} \, \dif \bm{x} \dif \bm{y} = D_{KL}(f(\bm{x},\bm{y})\|f(\bm{x})f(\bm{y}))
    \label{eq:mi}
\end{equation}
\noindent where $D_{KL}(P_X\|P_Y)$ is the Kullback--Leibler divergence of the distribution $P_X$ with respect to $P_Y$. In broad terms, MI is a measure of statistical dependence, as it reflects the discrepancy between the true joint probability model and the statistically independent model. Clearly, $I(\bm{X},\bm{Y}) = 0$ if and only if $\bm{X}$ and $\bm{Y}$ are independent. Since MI does not impose any assumption on the distribution of the random vectors or the form of relationship between them, it is a very general dependence measure \citep{ince2017statistical}.

Sometimes, the dependence between $\bm{X}$ and $\bm{Y}$ may be driven by another random vector $\bm{W}$. With linear dependencies, this can be captured by the partial correlation coefficient (PCC) since it measures the linear relationship between two variables after taking the effects of the conditioning vector out of both variables (see \citealp{christensen2002plane}). However, with nonlinear dependencies, the PCC may be inappropriate. Instead, the natural congruent of PCC in the mutual information framework is the conditional mutual information (CMI). By analogy with (\ref{eq:mi}), the CMI between $\bm{X}$ and $\bm{Y}$ given $\bm{W}$, denoted by $I(\bm{X},\bm{Y} \mid \bm{W})$, is given by
\begin{align*}
    I(\bm{X},\bm{Y} \mid \bm{W}) &= \iiint f(\bm{x},\bm{y},\bm{w}) \log \frac{f(\bm{w})f(\bm{x},\bm{y},\bm{w})}{f(\bm{x},\bm{w})f(\bm{y},\bm{w})} \, \dif \bm{x} \dif \bm{y} \dif \bm{w}.\\
    & = \mathbb{E}_{\bm{W}}\Big(D_{KL}\big(f(\bm{x},\bm{y} \mid \bm{W})\|f(\bm{x} \mid \bm{W})f(\bm{y} \mid \bm{W})\big)\Big)
\end{align*}
\noindent As CMI also takes its roots from the Kullback--Leiber divergence, it shares the same properties as MI with application to conditional independence (see \citealp{gray2011entropy}, for a more in-depth discussion). However, one limitation of MI and CMI for causal analysis is that both are symmetric measures, i.e., $I(\bm{X},\bm{Y}) = I(\bm{Y},\bm{X})$ and $I(\bm{X},\bm{Y} \mid \bm{W}) = I(\bm{Y},\bm{X} \mid \bm{W})$. Hence, they do not indicate any direction of dependence. 

To formulate an analogous information-theoretic causal inference metric for time series, \cite{schreiber2000measuring} developed the concept of transfer entropy. Transfer entropy (TE) from a series $X_t$ to another series $Y_t$, denoted by $TE(X \rightarrow Y;k,\ell)$, is defined as the CMI between $Y_t$ and $\bm{X}_{t-k}$ given $\bm{Y}_{t-\ell}$ where $\bm{X}_{t-k} = (X_{t-1},\ldots,X_{t-k})^{\top}$ and $\bm{Y}_{t-\ell} = (Y_{t-1},\dots,Y_{t-\ell})^{\top}$, for some selected time lags $k$ and $\ell$. Mathematically, with $y = y_t$, $\bm{y}' = (y_{t-1}, \ldots,y_{t-\ell})^{\top}$ and $\bm{x}' = (x_{t-k}, \ldots,x_{t-1})^{\top}$, we get
\begin{equation*}
    TE(X \rightarrow Y;k,\ell) = I(Y_t,\bm{X}_{t-k} \mid \bm{Y}_{t-\ell}) = \iiint f(y,\bm{y}',\bm{x}') \log \frac{f(\bm{y}')f(y,\bm{y}',\bm{x}')}{f(y,\bm{y}')f(\bm{y}',\bm{x}')} \,\dif y \dif \bm{x}' \dif \bm{y}'.
\end{equation*}
\noindent In other words, $TE(X \rightarrow Y;k,\ell)$ measures the impact of the lagged values $\bm{X}_{t-k}$ on the series $Y_t$ given its own history $\bm{Y}_{t-\ell}$ which is conceptually similar to GC. While TE is based on conditional distributions, GC looks at the improvement in the prediction variance, i.e., stating $X$ ``Granger-causes'' $Y$ if $\mbox{Var}(\hat{Y}_t \mid \bm{Y}_{t-\ell},\bm{X}_{t-k}) < \mbox{Var}(\hat{Y}_t \mid \bm{Y}_{t-\ell})$, where $\hat{Y}_t$ is the optimal predictor of $Y_t$ \citep{ding2006granger,bressler2011wiener}. Nonetheless, both frameworks infer the causal influence of one time series to another and \cite{barnett2009granger} showed the equivalence of TE and GC under the Gaussianity and linearity assumptions.

In this paper, we exploit the link between copula theory and information theory in order to develop an efficient approach for estimating TE in filtered EEG signals. By expressing the joint probability distribution with the corresponding copula density in Equation~(\ref{eq:cop2}) and some change of variables in the integration, \cite{ma2011mutual} showed that MI is related to copula entropy. We use the same technique for TE, viewed as CMI, to obtain the following:
\begin{align}
    \begin{split}
        TE(X \rightarrow Y;k,\ell) \; 
        & = \int_{[0,1]^{k+\ell+1}} c(u_y,\bm{u}_{\bm{y}'},\bm{u}_{\bm{x}'}) \log \frac{c(\bm{u}_{\bm{y}'})c(u_y,\bm{u}_{\bm{y}'},\bm{u}_{\bm{x}'})}{c(u_y,\bm{u}_{\bm{y}'})c(\bm{u}_{\bm{y}'},\bm{u}_{\bm{x}'})} \,\dif u_y \dif \bm{u}_{\bm{x}'} \dif \bm{u}_{\bm{y}'},
    \end{split}
    \label{eq:TE}
\end{align}
\noindent where $u_y = F(y_t)$, $\bm{u}_{\bm{y}'} = (F(y_{t-1}),\ldots,F(y_{t-\ell}))^{\top}$ and $\bm{u}_{\bm{x}'} = (F(x_{t-k}),\ldots,F(x_{t-1}))^{\top}$ with $F(\cdot)$ as the marginal distribution of the stationary processes $\{X_t\}$ and $\{Y_t\}$. Hence, a major advantage of our proposed estimation approach is that TE estimation requires three but simple independent steps, namely, marginal estimation, copula estimation and integration. The (nonparametric) empirical distribution function is a usual option to represent marginal distributions; alternatively, parametric approaches may also be used as discussed on Section~\ref{subchap:margins}. For estimating copula densities, common techniques may be used, e.g., empirical copulas based on ranks or some parametric multivariate copulas. The last integration step is straightforward, when $k$ and $\ell$ are not too large, since many numerical methods are available, e.g., Monte Carlo integration. Thus, TE, together with its copula representation, offers a general and appropriate framework for causal inference in neural signals with less restrictive assumptions than commonly used.

\section{Spectral Transfer Entropy}\label{chap:methodology}

We now develop a transfer-entropy measure for studying dependence between nodes (channels, voxels or regions of interest) in a brain network. Although classical TE can provide evidence of significant information flow between nodes in a brain network, one limitation is that it does not give any specific information on the oscillatory activities that are involved in this information transfer. Our goal is to develop a metric that captures both the magnitude and direction of information flow, say, from a theta-oscillation in one channel to the gamma-oscillation in another channel. Our solution is to construct TE in the frequency domain through a new spectral causal measure which we shall call the \textit{spectral transfer entropy (STE)}.

\subsection{Definition and General Strategy}

Let $\mathcal{X}_t$ and $\mathcal{Y}_t$ denote the EEG recording at time $t$ of two channels $\mathcal{X}$ and $\mathcal{Y}$, $t = 1,\ldots,T$. For $\Omega_1, \Omega_2 \in \{\delta,\theta,\alpha,\beta,\gamma\}$, denote by $Z^{\Omega_1}_{\mathcal{X},t}$ and $Z^{\Omega_2}_{\mathcal{Y},t}$ the latent oscillatory component of $\mathcal{X}_t$ and $\mathcal{Y}_t$ that are associated with the frequency bands $\Omega_1$ and $\Omega_2$, respectively. Suppose $\{\mathcal{X}_t\}$ and $\{\mathcal{Y}_t\}$ are stationary processes that represent mixtures of several oscillatory components. That is, $\mathcal{X}_t = g_{\mathcal{X}}(\bm{Z}_{\mathcal{X},t})$ and $\mathcal{Y}_t = g_{\mathcal{Y}}(\bm{Z}_{\mathcal{Y},t})$, where $g_{\mathcal{X}}(\cdot)$ and $g_{\mathcal{Y}}(\cdot)$ are ``mixing'' functions, $\bm{Z}_{\mathcal{X},t} = \big(Z^{\delta}_{\mathcal{X},t}, Z^{\theta}_{\mathcal{X},t}, Z^{\alpha}_{\mathcal{X},t}, Z^{\beta}_{\mathcal{X},t}, Z^{\gamma}_{\mathcal{X},t} \big)^{\top}$, and $\bm{Z}_{\mathcal{Y},t} = \big(Z^{\delta}_{\mathcal{Y},t}, Z^{\theta}_{\mathcal{Y},t}, Z^{\alpha}_{\mathcal{Y},t}, Z^{\beta}_{\mathcal{Y},t}, Z^{\gamma}_{\mathcal{Y},t} \big)^{\top}$. Specifically, $g_{\mathcal{X}}(\cdot)$ and $g_{\mathcal{Y}}(\cdot)$ are functions that combine the contribution of all latent oscillatory components into the observable time series $\{\mathcal{X}_t\}$ and $\{\mathcal{Y}_t\}$. For instance, a linear mixing function assumes the form $g(\bm{Z}) = \bm{a}^{\top} \bm{Z}$, where $\bm{a} = \{a_j\}^{d}_{j=1}$, $\bm{Z} = \{Z_j\}^{d}_{j=1}$ with $a_j > 0$ for all $j$ such that $\sum^{d}_{j=1} a_j = 1$. Moreover, we do not restrict any interaction between the oscillatory components from each frequency bands, e.g., the theta-oscillations of $\mathcal{X}_t$ may have causal influence on the theta-oscillations, as well as the gamma-oscillations, of $\mathcal{Y}_t$. Here, the main interest is to derive the causality structure across the different oscillations of $\mathcal{X}_t$ and $\mathcal{Y}_t$.

In the formulation of STE, since ``frequency \textit{band}-specific'' oscillations are known to smoothly fluctuate upwards and downwards, we argue that most of the information about the series are contained among the peaks and among the troughs, or equivalently, among the largest amplitudes (absolute values). Thus, instead of looking at the causality between two oscillatory processes based on every time point, we fix the attention only to the values where the series attain the maximum magnitudes. Define $\mathcal{M}^{\Omega_1}_{\mathcal{X},b} = \max \big(|Z^{\Omega_1}_{\mathcal{X},t}|; t  \in \{t_b + 1, \ldots, t_b + m\} \big)$ and $\mathcal{M}^{\Omega_2}_{\mathcal{Y},b} = \max \big(|Z^{\Omega_2}_{\mathcal{Y},t}| ; t  \in \{t_b + 1, \ldots, t_b + m\} \big)$ where $t_b$ is the time point preceding the $b$-th time block, i.e., $\mathcal{M}^{\Omega_1}_{\mathcal{X},b}$ and $\mathcal{M}^{\Omega_2}_{\mathcal{Y},b}$ represent the maximum magnitude of the oscillations $Z^{\Omega_1}_{\mathcal{X},t}$ and $Z^{\Omega_2}_{\mathcal{Y},t}$, respectively, over some time block of length $m$. Explicitly, through the STE, our objective is to infer the causal impact of $\mathcal{M}^{\Omega_1}_{\mathcal{X},b}$ on $\mathcal{M}^{\Omega_2}_{\mathcal{Y},b}$ (and vice versa).

We now define the spectral transfer entropy from an oscillatory component $Z^{\Omega_1}_{\mathcal{X},t}$ of $\mathcal{X}_t$ to the oscillatory component $Z^{\Omega_2}_{\mathcal{Y},t}$ of $\mathcal{Y}_t$, denoted by $STE_{\Omega_1,\Omega_2}(\mathcal{X} \rightarrow \mathcal{Y};k,\ell)$, as
\begin{equation}
    STE_{\Omega_1,\Omega_2}(\mathcal{X} \rightarrow \mathcal{Y};k,\ell) = TE(\mathcal{M}^{\Omega_1}_{\mathcal{X}} \rightarrow \mathcal{M}^{\Omega_2}_{\mathcal{Y}};k,\ell) = I(\mathcal{M}^{\Omega_2}_{\mathcal{Y},b},\bm{\mathcal{M}}^{\Omega_1}_{\mathcal{X},b-k} \mid \bm{\mathcal{M}}^{\Omega_2}_{\mathcal{Y},b-\ell}),
\end{equation}
\noindent where $\bm{\mathcal{M}}^{\Omega_1}_{\mathcal{X},b-k} = \big( \mathcal{M}^{\Omega_1}_{\mathcal{X},b-1}, \ldots, \mathcal{M}^{\Omega_1}_{\mathcal{X},b-k} \big)^{\top}$ and $\bm{\mathcal{M}}^{\Omega_1}_{\mathcal{Y},b-\ell} = \big( \mathcal{M}^{\Omega_1}_{\mathcal{Y},b-1}, \ldots, \mathcal{M}^{\Omega_1}_{\mathcal{Y},b-\ell} \big)^{\top}$. By its definition, STE enjoys all the nice properties of TE (and CMI). It quantifies the amount of information transferred from a band-specific oscillation of a series to an oscillation of another series. More importantly, having zero STE implies that one oscillatory component does not have a spectral causal influence to the other. A caveat, however, is that the temporal resolution of causality that the STE captures is at the level of the block maxima series and not the time scale of the original observed data. For example, if the EEG signals are recorded at a sampling rate of $128$ Hz (which is equivalent to having 128 observations per second) and the specified block size is $m = 32$, the causality is then defined to occur at the rate of about ``every quarter of a second''. Further discussion on the choice of $m$ and the temporal resolution of causality is included in Section~\ref{chap:numexp}. Nonetheless, STE bridges TE to the frequency domain in the context of band-specific oscillations.

\begin{figure}
	\centerline{
		\includegraphics[width=0.75\textwidth]{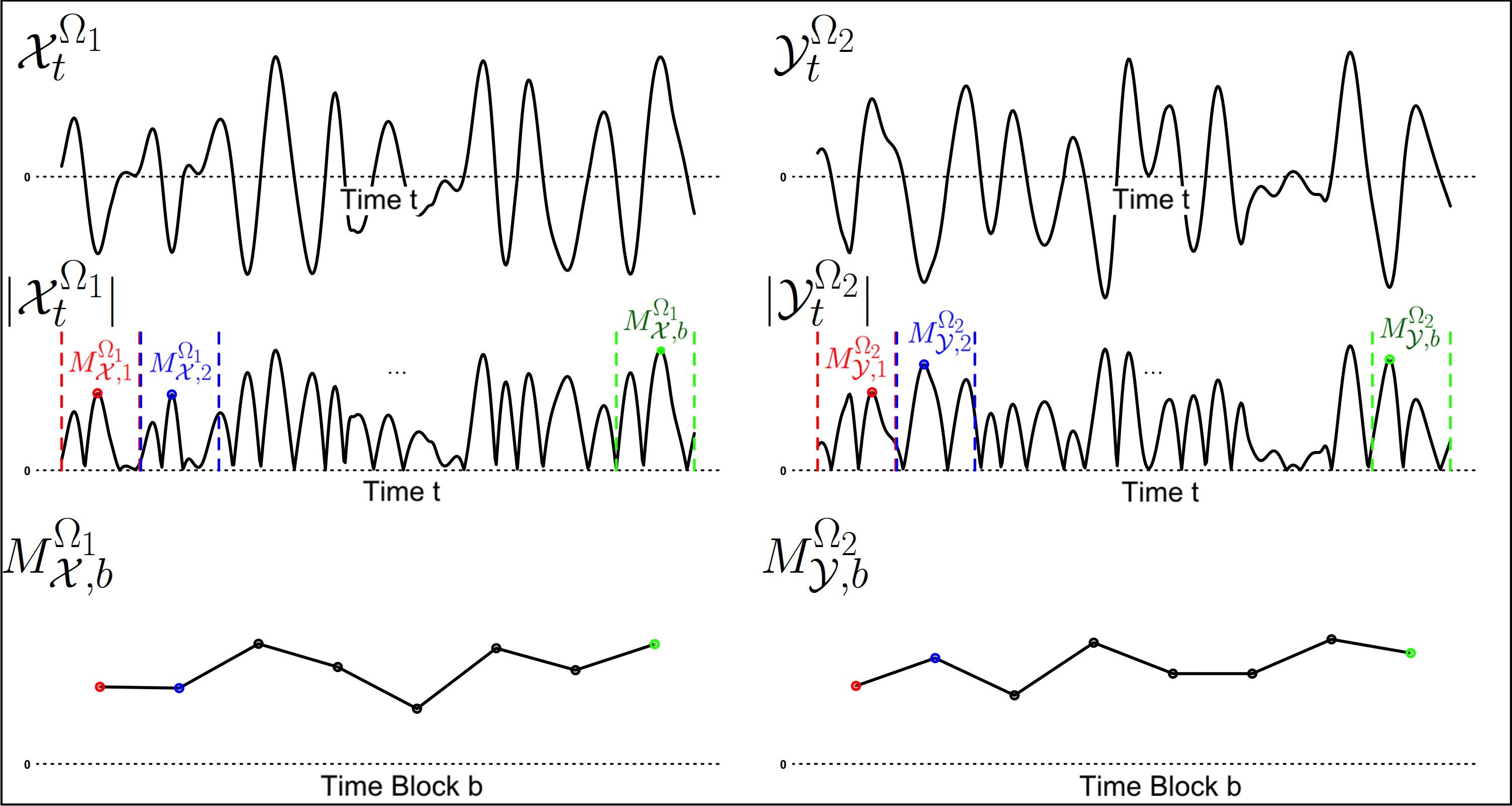}
	}
	\caption{Illustration for the filtered band-specific series $\mathcal{X}^{\Omega_1}_t$ and $\mathcal{Y}^{\Omega_2}_t$, their corresponding magnitudes $|\mathcal{X}^{\Omega_1}_t|$ and $|\mathcal{Y}^{\Omega_2}_t|$, and the computed non-overlapping block maxima series $M^{\Omega_1}_{\mathcal{X},b}$ and $M^{\Omega_2}_{\mathcal{Y},b}$.}
	\label{fig:OBM}
\end{figure}

Now, since $\mathcal{M}^{\Omega_1}_{\mathcal{X},b}$ and $\mathcal{M}^{\Omega_2}_{\mathcal{Y},b}$ are functions of $Z^{\Omega_1}_{\mathcal{X},t}$ and $Z^{\Omega_2}_{\mathcal{Y},t}$ which are unobservable latent processes, estimation of STE requires extraction of these oscillatory components from the observed signals. One way to do this is by filtering. The advantage of using the maximum magnitudes over time blocks is that, given the block size $m$ is large enough, it avoids the well-known issues caused by filtering on causal inference in the frequency domain (temporal dependence distortion and false extraction of spectral influence); see \cite{florin2010effect}, \cite{barnett2011behaviour} and \cite{seth2015granger}. Precisely, we perform the spectral causal inference based on STE as follows:
\begin{enumerate}
    \item For $\Omega_1, \Omega_2 \in \{\delta,\theta,\alpha,\beta,\gamma\}$, denote $\{\psi_j^{\Omega_1}\}$ and $\{\psi_j^{\Omega_2}\}$ to be the Butterworth (BW) filter coefficients with gain function concentrated on the bands $\Omega_1$ and $\Omega_2$, respectively. Define $\mathcal{X}^{\Omega_1}_t = \sum^{\infty}_{j=-\infty} \psi_j^{\Omega_1} \mathcal{X}_{t-j}$ and $\mathcal{Y}^{\Omega_2}_t = \sum^{\infty}_{j=-\infty} \psi_j^{\Omega_2} \mathcal{Y}_{t-j}$, which represent the extracted band-specific oscillations of $\mathcal{X}_t$ and $\mathcal{Y}_t$ for the five frequency bands.

    \item Let $M^{\Omega_1}_{\mathcal{X},b} = \max \big(|\mathcal{X}^{\Omega_1}_t| ; t \in \{(b-1)m + 1, \ldots, bm\} \big)$ and $M^{\Omega_2}_{\mathcal{Y},b} = \max \big(|\mathcal{Y}^{\Omega_2}_t| ;\\ t \in \{(b-1)m + 1, \ldots, bm\} \big)$ for $b \in \{1,\ldots,\lfloor\frac{T}{m}\rfloor\}$. That is, consider the maximum magnitude (absolute value) of the band-specific filtered series within non-overlapping time blocks of length $m$ (see Figure~\ref{fig:OBM}). 

    \item Serving as the analogue for $\mathcal{M}^{\Omega_1}_{\mathcal{X},b}$ and $\mathcal{M}^{\Omega_2}_{\mathcal{Y},b}$, estimate STE based on $M^{\Omega_1}_{\mathcal{X},b}$ and $M^{\Omega_2}_{\mathcal{Y},b}$ using the proposed estimation approach discussed below in Section~\ref{subchap:TEestimation} and perform significance testing based on the proposed resampling scheme, via the D-vine copula representation, outlined in Section~\ref{subchap:TEsigtest}.

    \item Since multiple connections are being tested simultaneously, apply a multiple comparison correction (e.g., the Benjamini--Hochberg (BH) method of \citealp{benjamini1995controlling} or the Bonferroni method of \citealp{bonferroni1935calcolo}) on the computed p-values to control for the false positive rate.
\end{enumerate}

We stress again that our approach uses the maximum magnitude within non-overlapping time blocks instead of the actual values of the filtered series across all time points. That is, we extract the information contained in the oscillatory processes through the highest amplitudes before measuring its causal impact onto one another. Hence, the proposed approach is robust to the inherent issues of filtering, in particular, artifacts in connectivity that is induced by the linear filter operator. Moreover, with a small finite number of frequency bands, controlling for false positive rates becomes feasible unlike for other frequency-specific methods, even when considering multiple channel pairs in the brain network. 

\subsection{Novel Copula-Based Estimation for STE}\label{subchap:TEestimation}

For the succeeding discussion, we re-label the series of block maxima for given frequency bands as $X_t$ and $Y_t$ instead of $M^{\Omega_1}_{\mathcal{X},b}$ and $M^{\Omega_2}_{\mathcal{Y},b}$. This is to avoid using more complex superscripts and subscripts, especially when defining lagged values of the block maxima series in the copula representation. Thus, the shift in notation implies that
\begin{equation*}
    STE_{\Omega_1,\Omega_2}(\mathcal{X} \rightarrow \mathcal{Y}; k,\ell) = TE(\mathcal{M}^{\Omega_1}_{\mathcal{X}} \rightarrow \mathcal{M}^{\Omega_2}_{\mathcal{Y}};k,\ell) \coloneqq TE(X \rightarrow Y;k,\ell).
\end{equation*}
\noindent Now, after extracting the band-specific series and computing the sequence of non-overlapping block maxima of the series' magnitudes, the estimation of STE is reduced to two main steps: fitting the marginal distributions for the series of block maxima (Section~\ref{subchap:margins}) and modeling the dependence structure associated with the joint distribution of the block maxima series (Section~\ref{subchap:vines}), before the integration (\ref{eq:TE}) is computed.

\subsubsection{Marginal Representations for Block Maxima Series}\label{subchap:margins}

Fitting the copula associated with $TE(X \rightarrow Y;k,\ell)$ requires first specifying the marginal distributions of $X_t$ and $Y_t$. This is an important task because incorrect assumptions on the margins may lead to a poor fit for the copula model, and thus, may result in false characterization of the dependence structure. There are several ways to estimate the margins $F(x_t)$ and $F(y_t)$. Since $X_t$ and $Y_t$ are series of block maxima, one appropriate choice is offered by the generalized extreme value (GEV) distribution. Suppose $\{W_j = |\mathcal{X}^{\Omega}_{t+j-1}|, \Omega \in \{\delta,\theta,\alpha,\beta,\gamma\}, j = 1, \ldots, m\}$ is a sequence of independent and identically distributed (IID) random variables and let $X_t = \max (W_1, \ldots, W_m) = \max \big(|\mathcal{X}^{\Omega}_{t}|, \ldots, |\mathcal{X}^{\Omega}_{t+m-1}|\big)$. By the Extremal Types Theorem \citep{fisher1928limiting,gnedenko1943distribution}, if there exist sequences of constants $a_m > 0$ and $b_m \in \mathbb{R}$, such that $\lim_{m \rightarrow \infty} \mbox{P}\left[a^{-1}_m \{\max(W_1, \ldots, W_m) - b_m\} \leq x\right] = G(x)$, for some non-degenerate distribution function $G$, then the limit $G$ has the form
\begin{equation}
    G(x) = \exp \left[-\Big\{1 + \xi \Big( \frac{x-\mu}{\sigma} \Big) \Big\}^{-1/\xi}_{+} \right],
      \label{eq:gev}
\end{equation}
\noindent where $a_{+} = \max(0,a)$. Equation~(\ref{eq:gev}) defines the GEV distribution with location parameter $\mu \in \mathbb{R}$, scale parameter $\sigma > 0$ and shape parameter $\xi \in \mathbb{R}$, and this family encompasses the three extreme value limit families: Fr\'{e}chet ($\xi > 0$), reversed Weibull ($\xi < 0$) and Gumbel ($\xi = 0$); see \citealp{coles2001introduction}. The Fr\'{e}chet distribution is heavy-tailed, the reversed Weibull distribution has an upper-bounded tail, and the Gumbel distribution is thin-tailed \citep{bali2003generalized}. For a review of univariate and multivariate models of extremes, see \cite{davison2015statistics}.

Because the sequence $\{\mathcal{X}^{\Omega}_{t}\}$ represents an oscillatory component of the observed EEG recordings, the IID assumption on $\{W_j\}$ is not appropriate. In fact, we assume the oscillatory components to be strictly stationary since frequency band-specific series, obtained from applying a linear filter (such as the BW filter) to stationary series, are also stationary \citep{brockwell2016introduction}. When $\{\mathcal{X}^{\Omega}_{t}\}$ is strictly stationary, is it straighforward to show that the absolute value sequence $\big\{|\mathcal{X}^{\Omega}_{t}|\big\}$ is also strictly stationary, but not IID. Fortunately, the extension of the Extremal Types Theorem to stationary sequences has been long established \citep{leadbetter1983extremes}. That is, under mild mixing conditions restricting long-range dependence, the limiting distribution for the renormalized maximum of the stationary sequence is still GEV, although the temporal dependence may modify the location and scale parameters \citep{davison2015statistics}. Hence, assuming GEV margins for each block maxima series $X_t$ and $Y_t$ offers a suitable representation of the marginal characteristics when the block size $m$ can be considered ``large''.

However, there are scenarios where the considered block size $m$ cannot be taken to be large, e.g., with limited data or when the causal analysis requires a higher temporal resolution. In such cases, another convenient option is simply to use the empirical cumulative distribution function (ECDF), i.e., $\hat{F}_n(x) = n^{-1}\sum^{n}_{i = 1} \text{I}(X_i \leq x)$, where $\text{I}(\cdot)$ is the indicator function. By the Glivenko--Cantelli Theorem (see, e.g., \citealp{wasserman2004all}), it converges to the true distribution as the sample size $n$ approaches to infinity, and hence, provides a consistent estimator for the marginal distributions. 

In the presence of non-stationarity, which is naturally expected when analyzing EEG signals, extensions of these two approaches may be employed by considering ``locally stationary'' time segments (see \citealp{ombao05,park2014estimating}; and \citealp{fiecas2016modeling}). That is, estimating the margins separately over portions of the observed data that behave almost like stationary series within. To be more precise, suppose we are interested in the margins of the block maxima series $\{X_t\}^{T}_{t=1}$. We may proceed as follows. Divide the horizon of $X_t$ into $Q$ locally stationary segments of length $M = \lfloor \frac{T}{Q} \rfloor$, let $\{X_t\}^{Mq}_{t = M(q-1) + 1}$ be the $q$-th time segment, and denote the marginal distribution of $X_t$ within the $q$-th segment by $F^q(x_t)$. Then, estimate $F^q(x_t)$ using a GEV distribution or using the ECDF. Thus, the estimated margins of $X_t$ over the entire time domain are defined as $\hat{F}(x_t) = \hat{F}^q(x_t), q = 1, \ldots, Q, t = M(q-1) + 1, \ldots Mq$. The same technique may be used to obtain $\hat{F}(y_t)$. By doing so, we ensure that the marginal characteristics are properly accounted for, and the observations are appropriately transformed to the uniform scale based on $\hat{F}(x_t)$ and $\hat{F}(y_t)$, before estimating their copula structure.

\subsubsection{Dependence Structure Through the D-vine Construction}\label{subchap:vines}

After taking into account the marginal characteristics of $X_t$ and $Y_t$ using appropriate distributional representations, the next step is to estimate the associated copula structure. While it may be natural to use multivariate max-stable distributions (i.e., extreme-value copulas) to model multivariate block maxima (see \cite{coles2001introduction} and \cite{gudendorf2012nonparametric} for definitions), we purposefully choose not to use such models to describe the dependence structure in our context because of several reasons: (i) it is difficult to model and estimate unstructured max-stable distributions in high dimensions \citep{davison2015statistics}; and (ii) recent research has shown that multivariate max-stable models are often too rigid in their dependence structure leading to poor fits in various applications \citep{huser2022advances}. In summary, we seek a flexible and fast framework to estimate the STE, and vine copulas meet these requirements. Hence, we assume a (simplified) vine copula model as defined in Equation~(\ref{eq:vine}) of Section~\ref{subchap:copula} and employ the sequential estimation approach proposed by \cite{aas2009pair} where the pair copulas that are involved in the decomposition of the $d$-dimensional copula are modeled separately. This allows for a more adaptable specification of the pairwise relationship between the variables as different copula families may be selected for each pair copula.

In addition, by strategically arranging the block maxima series and their respective lagged values as $(Y_t,Y_{t-1},\ldots,Y_{t-\ell},X_{t-k},\ldots,X_{t-1})^{\top}$ (equivalently, as $(u_y,\bm{u}_{\bm{y}'},\bm{u}_{\bm{x}'})^{\top}$ in (\ref{eq:TE})) in a D-vine structure, we further simplify the calculation of STE (and TE, in general). More concretely, because the $c(\bm{u}_{\bm{y}'})$, $c(u_y,\bm{u}_{\bm{y}'})$, and $c(\bm{u}_{\bm{y}'},\bm{u}_{\bm{x}'})$ are ``subsets'' of the full joint copula density, by expressing them into their corresponding D-vine decomposition, some common factors cancel out in the $\log$ component of $TE(X \rightarrow Y;k,\ell)$ (see Figure~\ref{fig:vinecopula}). This simplified form based on the D-vine naturally aligns with the definition of TE, i.e., the remaining pair copulas reflect the relationship between the current value of one series and the past values of another series given its own history. In general, for any $k$ and $\ell$, the innermost term in (\ref{eq:TE}) based on this simplified D-vine construction reduces to
\begin{equation}
    \log \frac{c(\bm{u}_{\bm{y}'})c(u_y,\bm{u}_{\bm{y}'},\bm{u}_{\bm{x}'})}{c(u_y,\bm{u}_{\bm{y}'})c(\bm{u}_{\bm{y}'},\bm{u}_{\bm{x}'})} = \sum^{k}_{j=1} \log c(F(y_t|\bm{y}_{t-\ell},\bm{x}_{t-(j+1)}),F(x_{t-j}|\bm{y}_{t-\ell},\bm{x}_{t-(j+1)})),
    \label{eq:vineTE}
\end{equation}
\noindent where $\bm{y}_{t-\ell} = (y_{t-1},\ldots,y_{t-\ell})^{\top}$ and $\bm{x}_{t-(j+1)} = (x_{t-(j+1)},x_{t-(j+2)},\ldots,x_{t-k})^{\top}$ with $\bm{x}_{t-(k+1)}$ removed from the conditioning set  (see Supplementary Material for the case when $k = \ell = 2$ and $k = \ell = 3$). As a result, the computation for $TE(X \rightarrow Y;k,\ell)$ only requires the remaining conditional bivariate copulas (not the full joint copula density) specified in Equation~(\ref{eq:vineTE}). With this, we estimate $STE_{\Omega_1,\Omega_2}(\mathcal{X} \rightarrow \mathcal{Y}; k,\ell)$ as follows:
\begin{figure}
	\centerline{
		\includegraphics[width=0.85\textwidth]{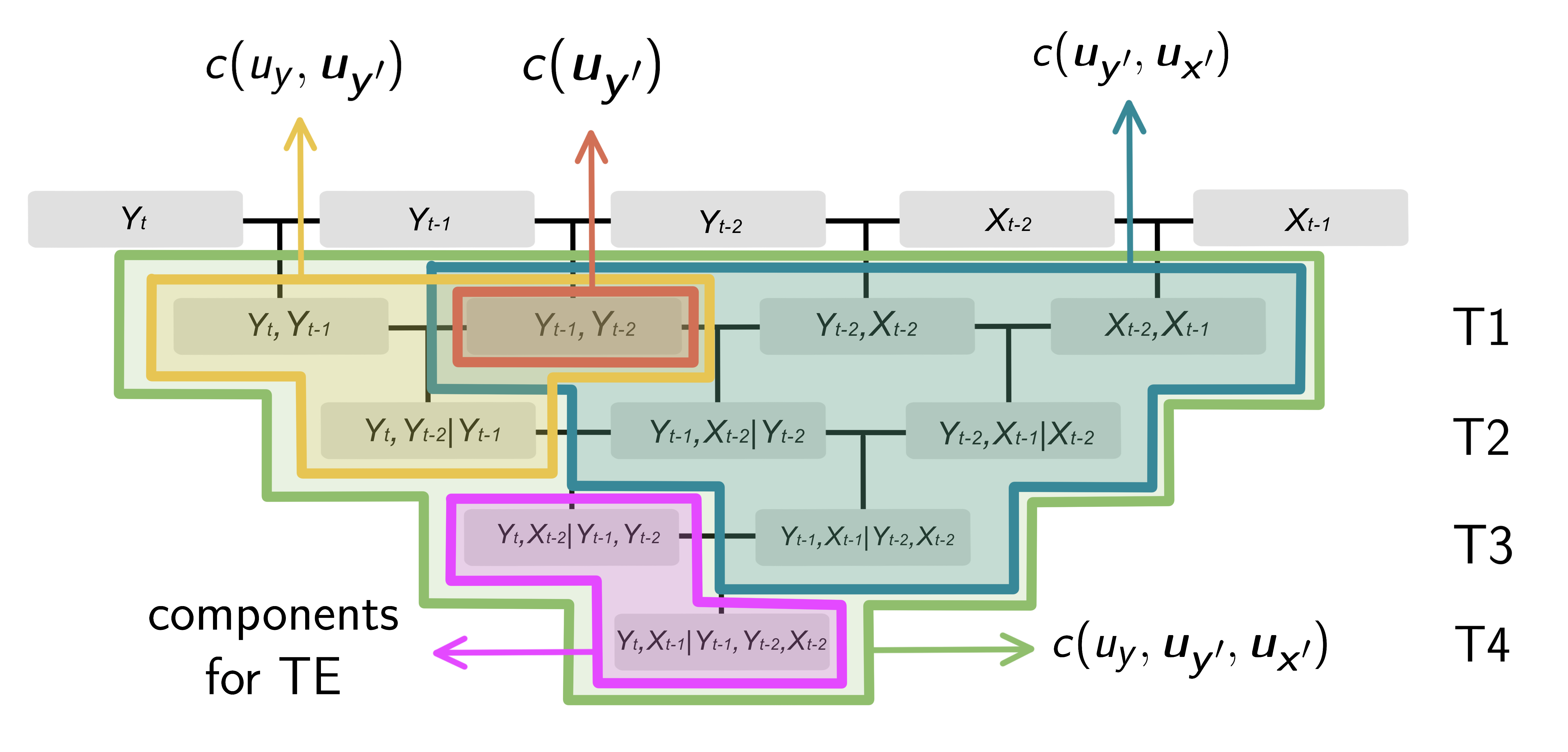}
	}
	\caption{Illustration for calculating $TE(X \rightarrow Y ; k = 2, \ell = 2)$ based on the strategically arranged D-vine copula structure of the series $Y_t$ and lagged series ($Y_{t-1},Y_{t-2},X_{t-1},X_{t-2}$).}
	\label{fig:vinecopula}
\end{figure}
\begin{enumerate}
    \item Fit individual marginal distributions for the block maxima series $X_t$ and $Y_t$ computed from the filtered series, and denote the fitted margins as $\hat{F}(x_t)$ and $\hat{F}(y_t)$.

    \item For pre-specified $k$ and $\ell$, arrange the variables in the D-vine structure and denote them by $(y_t,y_{t-1},\ldots,y_{t-\ell},x_{t-k},\ldots,x_{t-1},x_t)^{\top} \coloneqq (y,\bm{y}',\bm{x}',x)^\top$. Transform the variables to the uniform scale to get $(u_y,\bm{u}_{\bm{y}'},\bm{u}_{\bm{x}'},u_x)^{\top}$, where $u_y = \hat{F}(y_t)$, $\bm{u}_{\bm{y}'} = (\hat{F}(y_{t-1}),\ldots,\hat{F}(y_{t-\ell}))^{\top}$, $\bm{u}_{\bm{x}'} = (\hat{F}(x_{t-k}),\ldots,\hat{F}(x_{t-1}))^{\top}$, and $u_x = \hat{F}(x_t)$. Note that including $u_x$ is key to be able to estimate the causal interaction in both directions.
   
    \item Fit a D-vine copula model to $(u_y,\bm{u}_{\bm{y}'},\bm{u}_{\bm{x}'},u_x)^{\top}$ using the sequential estimation procedure developed by \cite{aas2009pair}. That is, for each pair copula in the decomposition, we select the ``best'' bivariate copula family among a chosen collection of copula models, based on some information criterion (e.g., modified Bayesian information criterion for vines (mBICv); see \citealp{nagler2019model}), and estimate the parameters of the chosen families. Choices for the bivariate copula selection may include the independence copula, elliptical copulas (e.g., Gaussian and Student's $t$), Archimedean copulas (e.g., Clayton, Frank, and Joe), and extreme-value copulas (e.g., Gumbel), or transformations thereof.
   
    \item Using the estimated D-vine copula, from Equation~(\ref{eq:vineTE}) and Monte Carlo integration, estimate $STE_{\Omega_1,\Omega_2}(\mathcal{X} \rightarrow \mathcal{Y}; k,\ell)$ using the estimator
    \vspace{-0.5mm}
    \begin{align*}
        \widehat{STE}_{\Omega_1,\Omega_2}&(\mathcal{X} \rightarrow \mathcal{Y}; k,\ell) = \widehat{TE}(X \rightarrow Y;k,\ell) \\
        & = (T^{*} - p)^{-1}\sum^{T^{*}}_{t = p + 1} \sum^{k}_{j=1} \log  \hat{c}(F(y^{*}_t|\bm{y}^{*}_{t-\ell},\bm{x}^{*}_{t-(j+1)}),F(x^{*}_{t-j}|\bm{y}^{*}_{t-\ell},\bm{x}^{*}_{t-(j+1)})),
    \end{align*}
    \noindent where $p = \max\{k,\ell\}$, $\{y^{*}_t, x^{*}_t\}^{T^{*}}_{t=1}$ are simulated observations from the fitted D-vine copula and the $\hat{c}(\cdot,\cdot)$s are the estimated pair copula densities from Step~(3). The number of observations $T^{*}$ is typically chosen to be large, e.g., $T^{*} = 10^4$.
\end{enumerate}

An advantage of our approach is its ability to provide estimates at the boundary point zero. Existing estimation procedures for TE requires ``shuffling'' to adjust the estimates from the bias due to finite sample effects \citep{marschinski2002analysing}. However, our method is robust to this problem. Whenever independent copulas are selected for the remaining conditional bivariate copula in Equation~(\ref{eq:vineTE}), the estimated TE attains an exact value of zero, thus removing the necessity for bias adjustment. Another advantage of our approach is that the estimate for the other causal direction, denoted by $\widehat{STE}_{\Omega_2,\Omega_1}(\mathcal{Y} \rightarrow \mathcal{X}; \ell,k) = \widehat{TE}(Y \rightarrow X;\ell,k)$, can be directly obtained from the same estimated D-vine copula model for $(u_y,\bm{u}_{\bm{y}'},\bm{u}_{\bm{x}'},u_x)^{\top}$. Hence, the new estimation scheme that we introduce provides a simple and efficient way for calculating STE in both causal directions based on a single fitted model, which reduces model selection bias.

\subsection{A Resampling Method for Testing Significance of Spectral Transfer Entropy}\label{subchap:TEsigtest}

Suppose we wish to test the following null hypothesis $H_0$,
\begin{align*}
    \begin{split}
    H_0: \;
    & STE_{\Omega_1,\Omega_2}(\mathcal{X} \rightarrow \mathcal{Y}; k,\ell) = TE(X \rightarrow Y;k,\ell) = 0\\
    \text{vs.~~}H_1: \;
    & STE_{\Omega_1,\Omega_2}(\mathcal{X} \rightarrow \mathcal{Y}; k,\ell) = TE(X \rightarrow Y;k,\ell) > 0.
    \end{split}
\end{align*}
Under the null hypothesis, $TE(X \rightarrow Y;k,\ell) = 0$ implies that the corresponding pair copulas $c(F(y_t|\bm{y}_{t-\ell},\bm{x}_{t-(j+1)}),F(x_{t-j}|\bm{y}_{t-\ell},\bm{x}_{t-(j+1)}))$, for $j = 1, \ldots, k$, are all equal to the independence copula. This enables simulating observations under $H_0$ from the fitted vine copula $\hat{C}(u_y,\bm{u}_{\bm{y}'},\bm{u}_{\bm{x}'},u_x)$ estimated from the original data.

\begin{figure}
	\centerline{
		\includegraphics[width=0.95\textwidth]{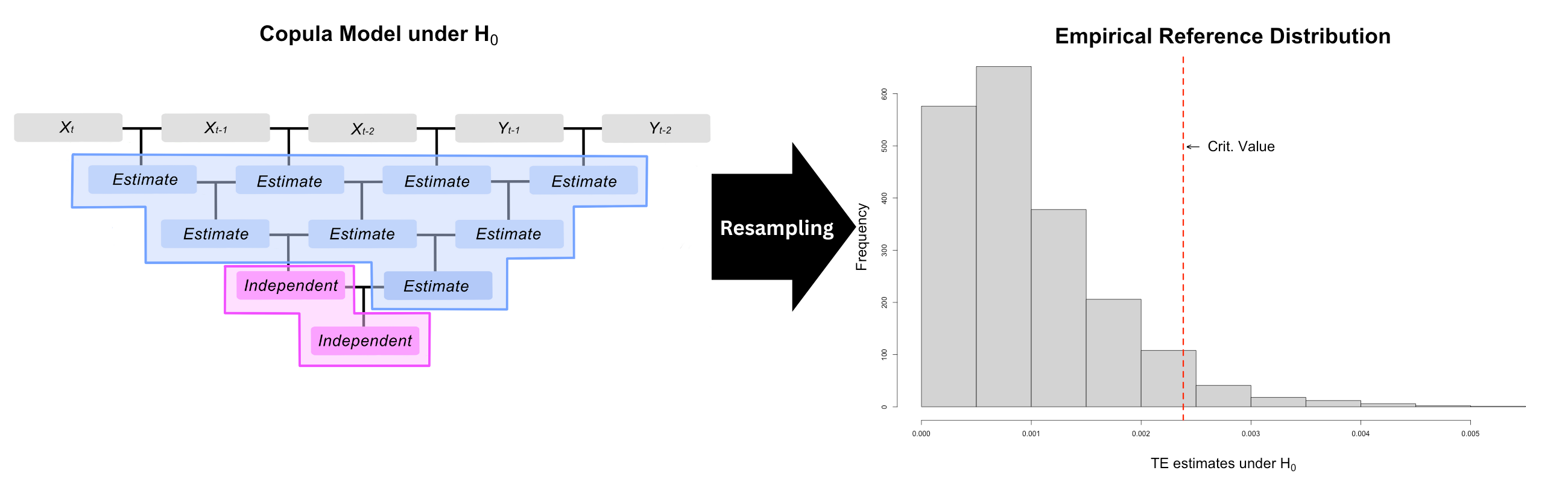}
	}
	\caption{The copula model under the null hypothesis and the generated empirical reference distribution for hypothesis testing.}
	\label{fig:resampTE}
\end{figure}

Denote the vine copula model under the null hypothesis as $C^{(0)}(u_y,\bm{u}_{\bm{y}'},\bm{u}_{\bm{x}'},u_x)$. The pair copulas involved in $C^{(0)}(\cdot)$ are the same bivariate copulas as those from the fitted model $\hat{C}(\cdot)$ except for the components that are specifically associated with $TE(X \rightarrow Y;k,\ell)$ which we replace by the independence copula (see Figure~\ref{fig:resampTE}). Since the independence copula density is given by $c(\cdot,\cdot) = 1$, the log component specified in (\ref{eq:vineTE}) reduces to $\sum^{k}_{j=1}\log (1) = 0$, resulting in an exact zero TE under $H_0$. Thus, new observations generated from $C^{(0)}(\cdot)$ preserve the dependence structure within each block maxima series $X$ and $Y$, but eliminate the information transfer from the past values $\bm{X}_{t-k}$ to $Y_t$. Repeatedly estimating TE from these generated observations then provides an empirical distribution of the estimator under the null hypothesis of no information transfer. Hence, we perform testing for significance of $TE(X \rightarrow Y;k,\ell)$ based on the outline below: 
\begin{enumerate}
    \item Generate new observations from the null vine copula model $C^{(0)}(\cdot)$ of the same size as the original data.

    \item Estimate $TE(X \rightarrow Y;k,\ell)$ from the new observations using the proposed copula-based estimation procedure and denote the estimates by $\widehat{TE}^{(r)}(X \rightarrow Y;k,\ell)$.

    \item Compute the p-value for testing the significance of the estimate $\widehat{TE}(X \rightarrow Y;k,\ell)$ as the relative frequency of the event $\{\widehat{TE}^{(r)}(X \rightarrow Y;k,\ell) \geq \widehat{TE}(X \rightarrow Y;k,\ell)\}$ among all resamples considered ($r = 1, \ldots, R$).

    \item Given a specified level of significance $\alpha \in (0,1)$, reject the null hypothesis if the p-value is less than $\alpha$. Otherwise, do not reject $H_0$.
\end{enumerate}

Another benefit offered by our procedure is the convenience in measuring uncertainties of the estimates. Since new observations can be easily generated from the null vine copula model, computing for significance levels (i.e., the associated p-values) using standard resampling methods is relatively fast even with a large number of replicates (to achieve an acceptable precision for multiple comparison adjustments). Moreover, simultaneously testing for both directions ($X \rightarrow Y$ and $Y \rightarrow X$) may be implemented from the same null copula model which greatly lessens the computational costs and model selection bias. Such feature is desirable in studying brain connectivity, especially when considering multiple pairs of nodes in a brain network while, at the same time, taking several oscillatory components into account.

\section{Numerical Experiments}\label{chap:numexp}

In this section, we explore the ability of our proposed causal metric for capturing information transfer in the frequency domain through realistic EEG simulations driven by the characterization that these signals are a function of oscillations at various frequency bands. Particularly, we provide evidence on the utility of STE in detecting significant and non-significant flow of information across different frequency oscillations, which may come in both linear and nonlinear forms. Moreover, we illustrate its robustness to the issues imposed by linear filtering, in comparison to the standard Wald test for Granger causality \citep{lutkepohl1992granger}, that is based on a fitted vector autoregressive (VAR) model. In addition, we outline a guide on how to specify tuning parameters of the STE metric, i.e., the choices for the block size $m$, and the number of considered lags $k$ and $\ell$, based on the findings from the simulation study.

To generate realistic EEG signals with known causal structure across different frequency oscillations, we borrow the concept of signal modulation from communication theory. Consider $Z_t = A_t \cos{(2\pi \omega t)}$ where $\cos{(2\pi \omega t)}$ is called the carrier signal at frequency $\omega \in (0,0.5)$ while $A_t$ is referred to as the modulating signal \citep{roder1931amplitude}. We adopt this formulation in the simulation of EEG signals and use the principle of amplitude modulation to induce causal relationships between the generated oscillatory components. That is, let $\varphi^{\Omega}_{\mathcal{X},t}$ and $\varphi^{\Omega}_{\mathcal{Y},t}$ be independent latent processes (carrier signals) such that their spectra concentrates in $\Omega \subset (0,0.5)$ and define $Z^{\Omega}_{\mathcal{X},t} = A^{\Omega}_{\mathcal{X},t} \varphi^{\Omega}_{\mathcal{X},t}$ and $Z^{\Omega}_{\mathcal{Y},t} = A^{\Omega}_{\mathcal{Y},t} \varphi^{\Omega}_{\mathcal{Y},t}$ where the modulating signals $A^{\Omega}_{\mathcal{X},t}$ and $A^{\Omega}_{\mathcal{Y},t}$ dictate the causal relationship between $Z^{\Omega}_{\mathcal{X},t}$ and $Z^{\Omega}_{\mathcal{Y},t}$. In communication science, information transfer occurs when a \textit{source} delivers a \textit{message} to a \textit{destination} \citep{bedford1962communication}. By considering channels $\mathcal{X}$ and $\mathcal{Y}$ as both sources and destinations of one another, the causal structure between their respective oscillatory components $Z^{\Omega}_{\mathcal{X},t}$ and $Z^{\Omega}_{\mathcal{Y},t}$ resembles the event of message delivery where the carrier signals $\varphi^{\Omega}_{\mathcal{X},t}$ and $\varphi^{\Omega}_{\mathcal{Y},t}$ serve as the pathway of information transfer while the modulating signals $A^{\Omega}_{\mathcal{X},t}$ and $A^{\Omega}_{\mathcal{Y},t}$ are the information being shared.

\begin{figure}
	\centerline{
		\includegraphics[width=0.95\textwidth]{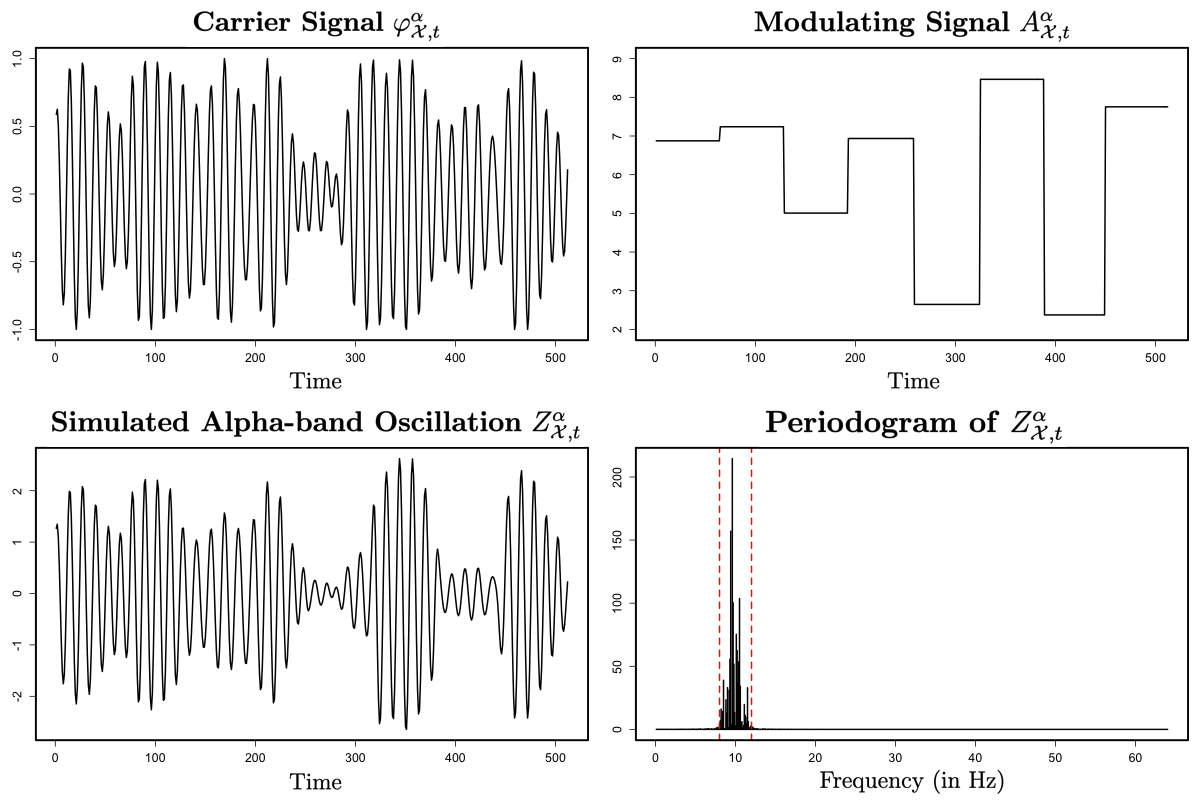}
	}
	\caption{Components of a simulated alpha-band oscillations; the carrier signal $\varphi^{\Omega}_{\mathcal{X},t}$ (top left), the modulating signal $A^{\Omega}_{\mathcal{X},t}$ (top right), the simulated oscillation $Z^{\Omega}_{\mathcal{X},t}$ (bottom left) and the periodogram of $Z^{\Omega}_{\mathcal{X},t}$ (bottom right).}
	\label{fig:sim_illus}
\end{figure}

In our simulations, we consider the following specifications for $A^{\Omega}_{\mathcal{X},t}$ and $A^{\Omega}_{\mathcal{Y},t}$. Let $\{(\vartheta_{\mathcal{X},b},\vartheta_{\mathcal{Y},b})^{\top}\}^{\infty}_{b = 1}$ follow a VAR process, and $\{t_{\mathcal{X},b}\}^{\infty}_{b = 1},\{t_{\mathcal{Y},b}\}^{\infty}_{b = 1} \sim F^{D}$ where $F^D$ is a discrete distribution defined on the integer set $\{\eta_{L},\eta_{L}+1, \ldots,\eta_{U}-1,\eta_{U}\}$ that has mean $\eta$ such that $\eta_L < \eta < \eta_U$, with $\eta_L,\eta,\eta_U \in \mathbb{Z}^{+}$. Define $T_{\mathcal{X},b} = \sum^{b}_{j=1}t_{\mathcal{X},j}$ and $T_{\mathcal{Y},b} = \sum^{b}_{j=1}t_{\mathcal{Y},j}$ for $b = 1, 2, \ldots$. Now, we let 
{\small
\begin{equation*}
    A^{\Omega}_{\mathcal{X},t} = \begin{cases}
        \vartheta_{\mathcal{X},1}, & t = 1, \ldots, T_{\mathcal{X},1}\\
        ~~~~\vdots \\
        \vartheta_{\mathcal{X},b}, & t = T_{\mathcal{X},b-1} +  1, \ldots, T_{\mathcal{X},b}\\
        ~~~~\vdots
    \end{cases} \text{~and~} A^{\Omega}_{\mathcal{Y},t} = \begin{cases}
        \vartheta_{\mathcal{Y},1}, & t = 1, \ldots, T_{\mathcal{Y},1}\\
        ~~~~\vdots \\
        \vartheta_{\mathcal{Y},b}, & t = T_{\mathcal{Y},b-1} +  1, \ldots, T_{\mathcal{Y},b}\\
        ~~~~\vdots
    \end{cases},
\end{equation*}
}
\noindent i.e., we consider $A^{\Omega}_{\mathcal{X},t}$ and $A^{\Omega}_{\mathcal{Y},t}$ to represent time-changing signal amplitudes, which take the form of a step function defined over random interval lengths given by $t_{\mathcal{X},b}$ and $t_{\mathcal{Y},b}$ (see Figure~\ref{fig:sim_illus} for an illustration). The advantage of such a formulation is that inducing causal relationships between $Z^{\Omega}_{\mathcal{X},t}$ and $Z^{\Omega}_{\mathcal{Y},t}$ is enabled by specifying an appropriate VAR process for $\{(\vartheta_{\mathcal{X},b},\vartheta_{\mathcal{Y},b})^{\top}\}^{B}_{b = 1}$, i.e., allowing for one-directional causal links ($Z^{\Omega}_{\mathcal{X}} \rightarrow Z^{\Omega}_{\mathcal{Y}}$ or $Z^{\Omega}_{\mathcal{Y}} \rightarrow Z^{\Omega}_{\mathcal{X}}$) or a two-way feedback ($Z^{\Omega}_{\mathcal{X}} \leftrightarrow Z^{\Omega}_{\mathcal{Y}}$). Moreover, we can generate the latent processes $\varphi^{\Omega}_{\mathcal{X},t}$ and $\varphi^{\Omega}_{\mathcal{Y},t}$ based on their respective autoregressive (AR) representations, following \cite{ombao2022spectral} and \cite{granados2022brain}, to mimic any of the five standard frequency bands: delta (0.5–4 Hz), theta (4–8 Hz), alpha (8–12 Hz), beta (12–30 Hz), and gamma (30–45 Hz). The technique outlined above ensures that the simulated oscillatory processes $Z^{\Omega}_{\mathcal{X},t}$ and $Z^{\Omega}_{\mathcal{Y},t}$ attain the desired causal relationship, while maintaining their respective spectra to be concentrated on the frequency band $\Omega$. Figure~\ref{fig:sim_illus} illustrates an example of the carrier signal $\varphi^{\Omega}_{\mathcal{X},t}$, the modulating signal $A^{\Omega}_{\mathcal{X},t}$, the oscillatory process $Z^{\Omega}_{\mathcal{X},t}$ and its corresponding periodogram.

\begin{figure}
	\centerline{
		\includegraphics[width=0.95\textwidth]{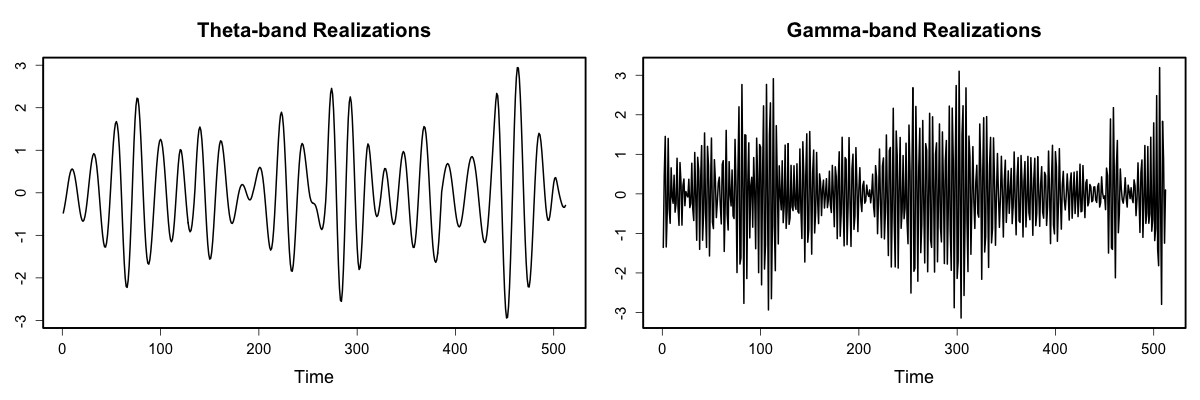}
	}
	\caption{Simulated theta-gamma coupling phenomenon.}
	\label{fig:thetagamma}
\end{figure}

Furthermore, by considering carrier signals with spectra concentrated on different frequency bands, e.g., $\varphi^{\Omega_1}_{\mathcal{X},t}$ and $\varphi^{\Omega_2}_{\mathcal{Y},t}$ where $\Omega_1, \Omega_2 \subset (0,0.5)$, generating cross-frequency band causal relationships across signals become possible under this framework. For instance, we simulate the theta-gamma ($\theta$--$\gamma$) coupling phenomenon by modulating the amplitude of the gamma latent process $Z^{\gamma}_{\mathcal{Y},t}$ with the magnitude of the theta latent process $Z^{\theta}_{\mathcal{X},t}$; see Figure~\ref{fig:thetagamma}. 

\begin{figure}
	\centerline{
		\includegraphics[width=0.35\textwidth]{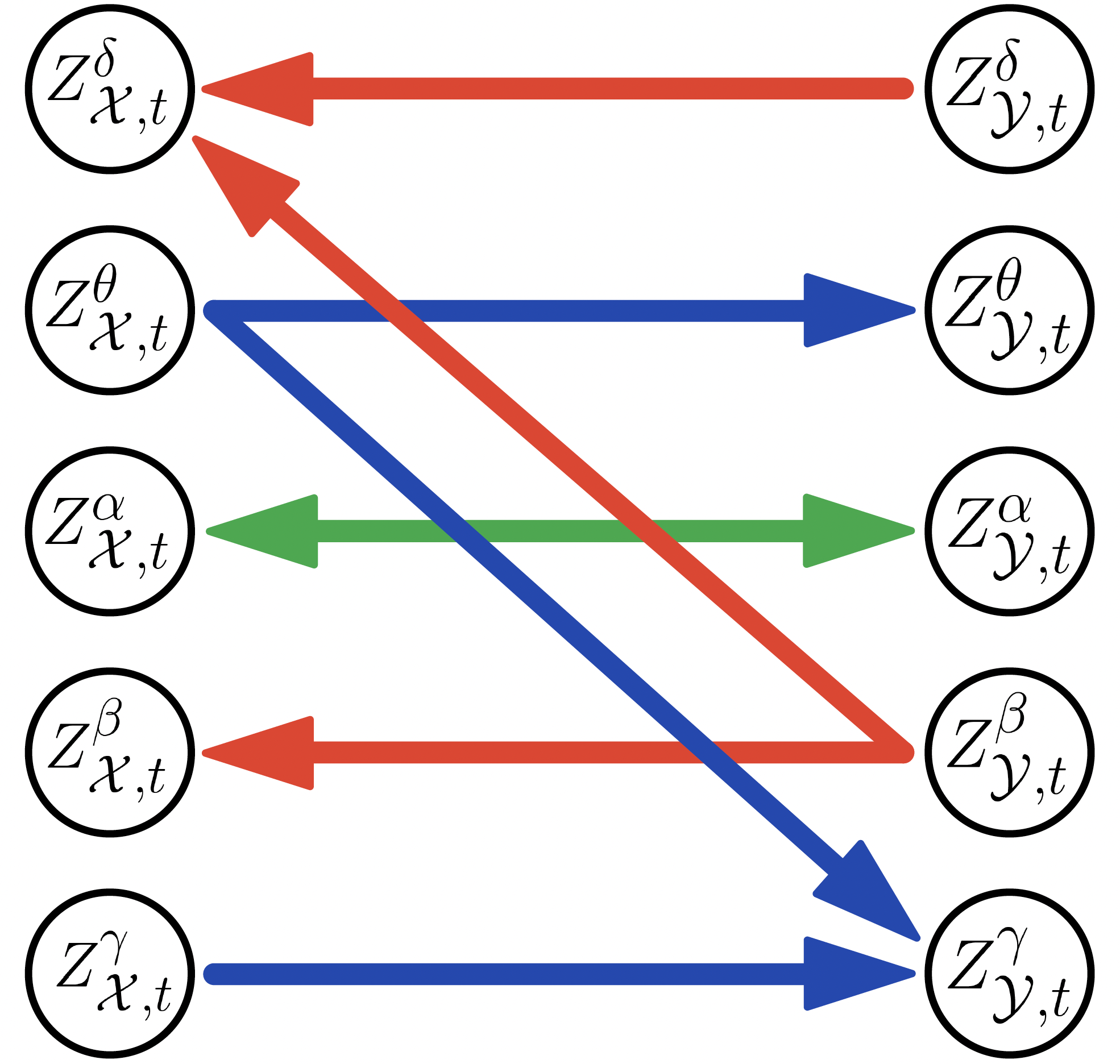}
	}
	\caption{Simulated directional links across the oscillatory components of $\mathcal{X}_t$ and $\mathcal{Y}_t$. Blue arrows indicate causal relationship from $\mathcal{X}_t$ to $\mathcal{Y}_t$, red arrows represent links from $\mathcal{Y}_t$ to $\mathcal{X}_t$, and green arrows denote two-way feedback between $\mathcal{X}_t$ and $\mathcal{Y}_t$.}
	\label{fig:sim_illus2}
\end{figure}

Denote $\{Z^{\Omega}_{\mathcal{X},t},Z^{\Omega}_{\mathcal{Y},t}\}$ to be the pair of simulated oscillatory processes coming from the same frequency band $\Omega$ where $\Omega \in \{\delta,\theta,\alpha,\beta,\gamma\}$. Figure~\ref{fig:sim_illus2} summarizes the induced causal structure, that we considered in our simulation experiment, between $Z^{\Omega}_{\mathcal{X},t}$ and $Z^{\Omega}_{\mathcal{Y},t}$ within the same and across different frequency bands. Details on the models used for simulation are reported in the Supplementary Material. With this, we construct the two series $\mathcal{X}_t$ and $\mathcal{Y}_t$ through linear mixing functions of these latent oscillatory processes, that is, $\mathcal{X}_t = g_{\mathcal{X}}(Z^{\delta}_{\mathcal{X},t},Z^{\theta}_{\mathcal{X},t},Z^{\alpha}_{\mathcal{X},t}, Z^{\beta}_{\mathcal{X},t},Z^{\gamma}_{\mathcal{X},t},W_{\mathcal{X},t})$ and $\mathcal{Y}_t = g_{\mathcal{Y}}(Z^{\delta}_{\mathcal{Y},t},Z^{\theta}_{\mathcal{Y},t},Z^{\alpha}_{\mathcal{Y},t},Z^{\beta}_{\mathcal{Y},t},Z^{\gamma}_{\mathcal{Y},t},W_{\mathcal{Y},t})$, where $g_{\mathcal{X}}(\cdot)$ and $g_{\mathcal{Y}}(\cdot)$ are linear functions with coefficients summing to one, and $W_{\mathcal{X},t},W_{\mathcal{Y},t} \overset{\mathrm{iid}}{\sim} N(0,1)$. To be more precise, all band-specific unit-variance latent oscillations are weighted equally in the linear mixture, and the coefficients are chosen such that the signal-to-noise ratio is around 95\%. This ensures that the BW filter can adequately extract the oscillations and the detected causal relationships are solely because of the proposed method---not artifacts of poor filtering. Testing the ability of the choice of filter to successfully extract the oscillatory components is outside the scope of this work. We emphasize that all simulated causal influences across the frequency bands originate from modulating signals whose relationships are linear by the nature of VAR processes; however, our construction yields non-linear causality patterns since the oscillations are defined as product of the modulating signal and the latent carrier signal. The novelty of STE is that it captures all (i.e., both linear and nonlinear) types of causal relationships as supported by the empirical evidence reported below.

To investigate the performance of the proposed measure, we consider several settings for the data generating process of $\mathcal{X}_t$ and $\mathcal{Y}_t$ to mimic scenarios that are typically encountered in practice. Precisely, we simulate the oscillatory components of $\mathcal{X}_t$ and $\mathcal{Y}_t$ (at sampling rate $s = 128$ Hz) assuming that $(\eta_L,\eta,\eta_U)^{\top} \in \{(29,32,35)^{\top},(61,64,67)^{\top}\}$. This choice of parameters implies that the simulated information transfer occurs at two possible rates, i.e., about  every quarter of a second (faster) or about every half a second (slower). Moreover, these signals are simulated for varying time series lengths corresponding to $N$ seconds worth of EEG recordings where $N \in \{15,30,60\}$. On the other hand, we explore different configurations for the tuning parameters of the STE metric across these scenarios. For each considered scenario, we estimated the STE measure based on the outlined D-vine copula approach with the ECDF for estimating the margins, and tested its significance using the proposed resampling scheme setting the block size as $m \in \{32,64\}$, and the number of lags as $k = \ell \coloneqq p \in \{1,2,3\}$. Numerical experiments are performed over $500$ replicates for all scenarios.

\begin{table}
\caption{Proportion of detected ``frequency band-specific'' causal links using the \textbf{WGC test} based on an estimated VAR($p$) model from $500$ simulation replicates (considering $0.05$ level of significance).}
\label{tab:sim_gc}
{\scriptsize
\begin{center}
\begin{tabular}{cc|cc|cc||cc|cc|cc}
\hline
\hline
 \multirow{2}{*}{$\mathbf{\mathcal{X}\rightarrow\mathcal{Y}}$} & \multirow{2}{*}{$\mathbf{N}$} & \multicolumn{2}{c|}{$\mathbf{\eta = 32}$} & \multicolumn{2}{c||}{$\mathbf{\eta = 64}$} & 
 \multirow{2}{*}{$\mathbf{\mathcal{Y}\rightarrow\mathcal{X}}$} & \multirow{2}{*}{$\mathbf{N}$} & \multicolumn{2}{c|}{$\mathbf{\eta = 32}$} & \multicolumn{2}{c}{$\mathbf{\eta = 64}$} \\
 \cline{3-4}\cline{5-6} \cline{9-10}\cline{11-12}
 &  & $\mathbf{p=2}$ & $\mathbf{p=5}$ & $\mathbf{p=2}$ & $\mathbf{p=5}$ & 
 &  & $\mathbf{p=2}$ & $\mathbf{p=5}$ & $\mathbf{p=2}$ & $\mathbf{p=5}$ \\
\hline
\hline

 \multirow{3}{*}{$\delta \rightarrow \delta$} & 15 & 0.728 & 0.942 & 0.708 & 0.902 &
 \multirow{3}{*}{$\delta \rightarrow \delta ~(*)$} & 15 & 0.708 & 0.918 & 0.720 & 0.908 \\
   & 30 & 0.746 & 0.930 & 0.714 & 0.920 & 
   & 30 & 0.734 & 0.930 & 0.714 & 0.912 \\
   & 60 & 0.710 & 0.904 & 0.706 & 0.916 &  
   & 60 & 0.708 & 0.898 & 0.718 & 0.918 \\
   \hline
   
 \multirow{3}{*}{$\theta \rightarrow \theta ~(*)$} & 15 & 0.698 & 0.942 & 0.720 & 0.948 &
 \multirow{3}{*}{$\theta \rightarrow \theta$} & 15 & 0.724 & 0.954 & 0.700 & 0.960 \\
   & 30 & 0.732 & 0.932 & 0.742 & 0.920 & 
   & 30 & 0.724 & 0.928 & 0.734 & 0.914 \\
   & 60 & 0.748 & 0.922 & 0.720 & 0.918 & 
   & 60 & 0.734 & 0.918 & 0.736 & 0.918 \\
   \hline
   
 \multirow{3}{*}{$\alpha \rightarrow \alpha ~(*)$} & 15 & 0.798 & 0.976 & 0.694 & 0.980 &
 \multirow{3}{*}{$\alpha \rightarrow \alpha ~(*)$} & 15 & 0.804 & 0.970 & 0.714 & 0.978 \\
   & 30 & 0.770 & 0.980 & 0.732 & 0.980 & 
   & 30 & 0.792 & 0.980 & 0.746 & 0.978 \\
   & 60 & 0.774 & 0.976 & 0.738 & 0.976 &
   & 60 & 0.776 & 0.968 & 0.720 & 0.968 \\
   \hline
   
 \multirow{3}{*}{$\beta \rightarrow \beta$} & 15 & 0.174 & 0.354 & 0.222 & 0.390 &
 \multirow{3}{*}{$\beta \rightarrow \beta ~(*)$} & 15 & 0.158 & 0.360 & 0.202 & 0.374 \\
   & 30 & 0.176 & 0.394 & 0.184 & 0.364 &  
   & 30 & 0.168 & 0.402 & 0.172 & 0.354 \\
   & 60 & 0.184 & 0.358 & 0.180 & 0.346 & 
   & 60 & 0.182 & 0.382 & 0.204 & 0.360 \\
   \hline
   
 \multirow{3}{*}{$\gamma \rightarrow \gamma ~(*)$} & 15 & 0.298 & 0.614 & 0.296 & 0.650 &
 \multirow{3}{*}{$\gamma \rightarrow \gamma$} & 15 & 0.294 & 0.610 & 0.306 & 0.628 \\
   & 30 & 0.340 & 0.614 & 0.302 & 0.650 & 
   & 30 & 0.344 & 0.642 & 0.312 & 0.654 \\
   & 60 & 0.324 & 0.624 & 0.278 & 0.596 & 
   & 60 & 0.312 & 0.622 & 0.288 & 0.636 \\
   \hline
   
 \multirow{3}{*}{$\delta \rightarrow \beta$} & 15 & 0.000 & 0.000 & 0.000 & 0.000 & 
 \multirow{3}{*}{$\beta \rightarrow \delta ~(*)$} & 15 & 0.000 & 0.000 & 0.000 & 0.000 \\
   & 30 & 0.000 & 0.000 & 0.000 & 0.000 &
   & 30 & 0.000 & 0.000 & 0.000 & 0.000 \\
   & 60 & 0.000 & 0.000 & 0.000 & 0.000 & 
   & 60 & 0.000 & 0.000 & 0.000 & 0.000 \\
   \hline
   
 \multirow{3}{*}{$\theta \rightarrow \gamma ~(*)$} & 15 & 0.000 & 0.000 & 0.000 & 0.000 & 
 \multirow{3}{*}{$\gamma \rightarrow \theta$} & 15 & 0.000 & 0.000 & 0.000 & 0.000 \\
   & 30 & 0.000 & 0.000 & 0.000 & 0.000 &   
   & 30 & 0.000 & 0.000 & 0.000 & 0.000 \\
   & 60 & 0.000 & 0.000 & 0.000 & 0.000 &  
   & 60 & 0.000 & 0.000 & 0.000 & 0.000 \\

\hline
\hline

\end{tabular}
\end{center}
}
\flushleft{\vspace{-1.5mm}\footnotesize{~~The simulated significant ``frequency \textit{band}-specific'' causal links are marked with $(*)$.}}
\end{table}

{\setlength{\tabcolsep}{2.5pt}
\begin{table}
\caption{Proportion of detected ``frequency band-specific'' causal links from signals with simulated information transfer rate of about every quarter of a second ($\mathbf{\eta = 32}$) based on the \textbf{proposed STE metric} from $500$ simulation replicates (considering $0.05$ level of significance).}
\label{tab:sim_r1}
{\scriptsize
\begin{center}
\begin{tabular}{cc|ccc|ccc||cc|ccc|ccc}
\hline
\hline
 \multirow{2}{*}{$\mathbf{\mathcal{X}\rightarrow\mathcal{Y}}$} & \multirow{2}{*}{$\mathbf{N}$} & \multicolumn{3}{c|}{$\mathbf{m = 32}$} & \multicolumn{3}{c||}{$\mathbf{m = 64}$} & 
 \multirow{2}{*}{$\mathbf{\mathcal{Y}\rightarrow\mathcal{X}}$} & \multirow{2}{*}{$\mathbf{N}$} & \multicolumn{3}{c|}{$\mathbf{m = 32}$} & \multicolumn{3}{c}{$\mathbf{m = 64}$} \\
 \cline{3-5}\cline{6-8} \cline{11-13}\cline{14-16}
 &  & $\mathbf{p=1}$ & $\mathbf{p=2}$ & $\mathbf{p=3}$ & $\mathbf{p=1}$ & $\mathbf{p=2}$ & $\mathbf{p=3}$ & 
 &  & $\mathbf{p=1}$ & $\mathbf{p=2}$ & $\mathbf{p=3}$ & $\mathbf{p=1}$ & $\mathbf{p=2}$ & $\mathbf{p=3}$ \\
\hline
\hline

  \multirow{3}{*}{$\delta \rightarrow \delta$} & 15 & 0.046 & 0.038 & 0.030 & 0.054 & 0.052 & 0.054 & 
  \multirow{3}{*}{$\delta \rightarrow \delta ~(*)$} & 15 & 0.278 & 0.247 & 0.208 & 0.134 & 0.106 & 0.102 \\
   & 30 & 0.038 & 0.034 & 0.038 & 0.032 & 0.038 & 0.048 &   & 30 & 0.386 & 0.356 & 0.284 & 0.234 & 0.114 & 0.148 \\
   & 60 & 0.036 & 0.048 & 0.042 & 0.032 & 0.034 & 0.036 &   & 60 & 0.572 & 0.542 & 0.510 & 0.346 & 0.290 & 0.172 \\
   \hline
   
 \multirow{3}{*}{$\theta \rightarrow \theta ~(*)$} & 15 & 0.814 & 0.748 & 0.678 & 0.376 & 0.252 & 0.188 &
 \multirow{3}{*}{$\theta \rightarrow \theta$} & 15 & 0.056 & 0.033 & 0.040 & 0.050 & 0.054 & 0.062 \\
   & 30 & 0.960 & 0.912 & 0.868 & 0.558 & 0.460 & 0.414 & 
   & 30 & 0.060 & 0.042 & 0.048 & 0.038 & 0.036 & 0.040 \\
   & 60 & 0.976 & 0.972 & 0.956 & 0.754 & 0.670 & 0.606 & 
   & 60 & 0.064 & 0.032 & 0.048 & 0.046 & 0.050 & 0.040 \\
   \hline
   
 \multirow{3}{*}{$\alpha \rightarrow \alpha ~(*)$} & 15 & 0.580 & 0.432 & 0.388 & 0.242 & 0.162 & 0.142 &
 \multirow{3}{*}{$\alpha \rightarrow \alpha ~(*)$} & 15 & 0.680 & 0.499 & 0.416 & 0.338 & 0.244 & 0.182 \\
   & 30 & 0.898 & 0.740 & 0.616 & 0.512 & 0.356 & 0.298 & 
   & 30 & 0.918 & 0.794 & 0.738 & 0.624 & 0.414 & 0.418 \\
   & 60 & 0.980 & 0.962 & 0.882 & 0.794 & 0.626 & 0.562 & 
   & 60 & 0.992 & 0.986 & 0.936 & 0.930 & 0.742 & 0.714 \\
   \hline
   
 \multirow{3}{*}{$\beta \rightarrow \beta$} & 15 & 0.036 & 0.031 & 0.034 & 0.050 & 0.044 & 0.048 &
 \multirow{3}{*}{$\beta \rightarrow \beta ~(*)$} & 15 & 0.994 & 0.993 & 0.980 & 0.514 & 0.338 & 0.318 \\
   & 30 & 0.034 & 0.030 & 0.036 & 0.044 & 0.066 & 0.034 & 
   & 30 & 0.998 & 0.994 & 0.996 & 0.724 & 0.608 & 0.526 \\
   & 60 & 0.070 & 0.056 & 0.080 & 0.040 & 0.036 & 0.038 & 
   & 60 & 0.988 & 0.990 & 0.992 & 0.866 & 0.800 & 0.750 \\
   \hline
   
 \multirow{3}{*}{$\gamma \rightarrow \gamma ~(*)$} & 15 & 0.936 & 0.911 & 0.848 & 0.338 & 0.192 & 0.158 &
 \multirow{3}{*}{$\gamma \rightarrow \gamma$} & 15 & 0.042 & 0.047 & 0.034 & 0.056 & 0.050 & 0.036 \\
   & 30 & 0.976 & 0.982 & 0.970 & 0.490 & 0.392 & 0.354 & 
   & 30 & 0.038 & 0.040 & 0.036 & 0.030 & 0.048 & 0.036 \\
   & 60 & 0.974 & 0.970 & 0.978 & 0.688 & 0.584 & 0.562 & 
   & 60 & 0.052 & 0.050 & 0.040 & 0.050 & 0.032 & 0.034 \\
   \hline
   
 \multirow{3}{*}{$\delta \rightarrow \beta$} & 15 & 0.036 & 0.029 & 0.052 & 0.054 & 0.056 & 0.062 & 
 \multirow{3}{*}{$\beta \rightarrow \delta ~(*)$} & 15 & 0.574 & 0.457 & 0.402 & 0.162 & 0.138 & 0.100 \\
   & 30 & 0.038 & 0.042 & 0.038 & 0.046 & 0.046 & 0.044 & 
   & 30 & 0.762 & 0.708 & 0.624 & 0.292 & 0.244 & 0.166 \\
   & 60 & 0.044 & 0.050 & 0.038 & 0.046 & 0.036 & 0.060 & 
   & 60 & 0.878 & 0.868 & 0.808 & 0.484 & 0.374 & 0.324 \\
   \hline
   
 \multirow{3}{*}{$\theta \rightarrow \gamma ~(*)$} & 15 & 0.772 & 0.702 & 0.614 & 0.248 & 0.174 & 0.112 & 
 \multirow{3}{*}{$\gamma \rightarrow \theta$} & 15 & 0.064 & 0.045 & 0.040 & 0.062 & 0.046 & 0.042 \\
   & 30 & 0.912 & 0.878 & 0.858 & 0.388 & 0.346 & 0.284 & 
   & 30 & 0.058 & 0.046 & 0.040 & 0.024 & 0.028 & 0.046 \\
   & 60 & 0.970 & 0.950 & 0.938 & 0.620 & 0.550 & 0.482 & 
   & 60 & 0.054 & 0.044 & 0.034 & 0.032 & 0.030 & 0.042 \\

\hline
\hline

\end{tabular}
\end{center}
}
\flushleft{\vspace{-1.5mm}\footnotesize{~~The simulated significant ``frequency \textit{band}-specific'' causal links are marked with $(*)$.}}
\end{table}
}

{\setlength{\tabcolsep}{2.5pt}
\begin{table}
\caption{Proportion of detected ``frequency band-specific'' causal links from signals with simulated information transfer rate of about every half a second ($\mathbf{\eta = 64}$) based on the \textbf{proposed STE metric} from $500$ simulation replicates (considering $0.05$ level of significance).}
\label{tab:sim_r2}
{\scriptsize
\begin{center}
\begin{tabular}{cc|ccc|ccc||cc|ccc|ccc}
\hline
\hline
 \multirow{2}{*}{$\mathbf{\mathcal{X}\rightarrow\mathcal{Y}}$} & \multirow{2}{*}{$\mathbf{N}$} & \multicolumn{3}{c|}{$\mathbf{m = 32}$} & \multicolumn{3}{c||}{$\mathbf{m = 64}$} & 
 \multirow{2}{*}{$\mathbf{\mathcal{Y}\rightarrow\mathcal{X}}$} & \multirow{2}{*}{$\mathbf{N}$} & \multicolumn{3}{c|}{$\mathbf{m = 32}$} & \multicolumn{3}{c}{$\mathbf{m = 64}$} \\
 \cline{3-5}\cline{6-8} \cline{11-13}\cline{14-16}
 &  & $\mathbf{p=1}$ & $\mathbf{p=2}$ & $\mathbf{p=3}$ & $\mathbf{p=1}$ & $\mathbf{p=2}$ & $\mathbf{p=3}$ & 
 &  & $\mathbf{p=1}$ & $\mathbf{p=2}$ & $\mathbf{p=3}$ & $\mathbf{p=1}$ & $\mathbf{p=2}$ & $\mathbf{p=3}$ \\
\hline
\hline
 \multirow{3}{*}{$\delta \rightarrow \delta$} & 15 & 0.060 & 0.040 & 0.050 & 0.064 & 0.060 & 0.062 & 
 \multirow{3}{*}{$\delta \rightarrow \delta ~(*)$} & 15 & 0.306 & 0.426 & 0.391 & 0.590 & 0.398 & 0.340 \\
   & 30 & 0.030 & 0.036 & 0.034 & 0.036 & 0.042 & 0.040 &  
   & 30 & 0.516 & 0.686 & 0.644 & 0.832 & 0.760 & 0.654 \\
   & 60 & 0.020 & 0.032 & 0.034 & 0.042 & 0.034 & 0.024 &  
   & 60 & 0.764 & 0.948 & 0.896 & 0.968 & 0.932 & 0.916 \\
   \hline
   
 \multirow{3}{*}{$\theta \rightarrow \theta ~(*)$} & 15 & 0.500 & 0.788 & 0.775 & 0.932 & 0.854 & 0.790 & 
 \multirow{3}{*}{$\theta \rightarrow \theta$} & 15 & 0.072 & 0.036 & 0.046 & 0.056 & 0.048 & 0.068 \\
   & 30 & 0.770 & 0.972 & 0.982 & 0.998 & 0.990 & 0.980 & 
   & 30 & 0.054 & 0.042 & 0.038 & 0.054 & 0.022 & 0.046 \\
   & 60 & 0.918 & 1.000 & 1.000 & 1.000 & 1.000 & 0.998 & 
   & 60 & 0.092 & 0.048 & 0.038 & 0.034 & 0.026 & 0.030 \\
   \hline
   
 \multirow{3}{*}{$\alpha \rightarrow \alpha ~(*)$} & 15 & 0.360 & 0.704 & 0.411 & 0.816 & 0.434 & 0.396 &
 \multirow{3}{*}{$\alpha \rightarrow \alpha ~(*)$} & 15 & 0.350 & 0.684 & 0.424 & 0.820 & 0.444 & 0.414 \\
   & 30 & 0.654 & 0.958 & 0.798 & 0.988 & 0.788 & 0.748 &  
   & 30 & 0.652 & 0.962 & 0.844 & 0.988 & 0.776 & 0.772 \\
   & 60 & 0.824 & 0.998 & 0.986 & 0.998 & 0.960 & 0.944 & 
   & 60 & 0.860 & 1.000 & 0.992 & 0.998 & 0.976 & 0.966 \\
   \hline
   
 \multirow{3}{*}{$\beta \rightarrow \beta$} & 15 & 0.028 & 0.034 & 0.036 & 0.052 & 0.050 & 0.050 &
 \multirow{3}{*}{$\beta \rightarrow \beta ~(*)$} & 15 & 0.656 & 0.984 & 0.977 & 0.992 & 0.994 & 0.986 \\
   & 30 & 0.044 & 0.034 & 0.038 & 0.054 & 0.034 & 0.050 & 
   & 30 & 0.896 & 0.998 & 1.000 & 1.000 & 1.000 & 1.000 \\
   & 60 & 0.024 & 0.048 & 0.044 & 0.046 & 0.036 & 0.042 & 
   & 60 & 0.960 & 1.000 & 1.000 & 1.000 & 1.000 & 1.000 \\
   \hline
   
 \multirow{3}{*}{$\gamma \rightarrow \gamma ~(*)$} & 15 & 0.362 & 0.780 & 0.725 & 0.918 & 0.786 & 0.712 &
 \multirow{3}{*}{$\gamma \rightarrow \gamma$} & 15 & 0.040 & 0.042 & 0.020 & 0.040 & 0.038 & 0.054 \\
   & 30 & 0.622 & 0.954 & 0.974 & 0.994 & 0.982 & 0.962 & 
   & 30 & 0.056 & 0.040 & 0.060 & 0.038 & 0.044 & 0.042 \\
   & 60 & 0.774 & 1.000 & 1.000 & 1.000 & 1.000 & 1.000 &  
   & 60 & 0.042 & 0.042 & 0.040 & 0.036 & 0.030 & 0.058 \\
   \hline
   
 \multirow{3}{*}{$\delta \rightarrow \beta$} & 15 & 0.046 & 0.044 & 0.046 & 0.050 & 0.038 & 0.040 & 
 \multirow{3}{*}{$\beta \rightarrow \delta ~(*)$} & 15 & 0.336 & 0.560 & 0.540 & 0.672 & 0.584 & 0.462 \\
   & 30 & 0.040 & 0.046 & 0.032 & 0.042 & 0.036 & 0.046 & 
   & 30 & 0.640 & 0.822 & 0.808 & 0.942 & 0.892 & 0.822 \\
   & 60 & 0.042 & 0.032 & 0.028 & 0.048 & 0.034 & 0.034 &  
   & 60 & 0.808 & 0.982 & 0.982 & 0.992 & 0.992 & 0.970 \\
   \hline
   
 \multirow{3}{*}{$\theta \rightarrow \gamma ~(*)$} & 15 & 0.354 & 0.612 & 0.609 & 0.786 & 0.680 & 0.548 & 
 \multirow{3}{*}{$\gamma \rightarrow \theta$} & 15 & 0.058 & 0.044 & 0.040 & 0.040 & 0.054 & 0.052 \\
   & 30 & 0.590 & 0.912 & 0.862 & 0.968 & 0.954 & 0.884 & 
   & 30 & 0.046 & 0.022 & 0.030 & 0.046 & 0.036 & 0.048 \\
   & 60 & 0.816 & 0.980 & 0.994 & 1.000 & 0.998 & 0.996 & 
   & 60 & 0.072 & 0.042 & 0.036 & 0.034 & 0.028 & 0.050 \\

\hline
\hline

\end{tabular}
\end{center}
}
\flushleft{\vspace{-1.5mm}\footnotesize{~~The simulated significant ``frequency \textit{band}-specific'' causal links are marked with $(*)$.}}
\end{table}
}

The problem when investigating causality from ``frequency band''-filtered signals is that linear filtering (e.g., using BW filters) induces a smooth oscillatory behavior on the filtered signals. This introduces false artifacts of temporal dependence resulting in the failure of standard causality frameworks (e.g., Granger causality) to appropriately capture directed spectral influence. We illustrate this issue by performing the standard Wald test for Granger causality (WGC) based on an estimated VAR($p$) model, where $p \in \{2,5\}$, fitted to the band-specific filtered series and highlight that our proposed STE measure is robust to this inherent issue of linear filtering. Tables~\ref{tab:sim_gc}--\ref{tab:sim_r2} summarize the proportion of detected significant causal links based on WGC and STE for some selected pairs of frequency bands (considering a $0.05$ level of significance) across the scenarios considered. From Table~\ref{tab:sim_gc}, we clearly see how the WGC framework suffers from the effects of filtering. When dealing with causality from the same frequency band, performing WGC on the filtered signals causes the method to detect significant information transfer for both directions even when only one directional link is significant. For example, although we simulated $Z^{\theta}_{\mathcal{X}} \rightarrow Z^{\theta}_{\mathcal{Y}}$ and $Z^{\theta}_{\mathcal{Y}} \not\rightarrow Z^{\theta}_{\mathcal{X}}$, WGC finds these two causal links to be significant leading to very high false positive rates. This false detection becomes more severe when a VAR model with larger lag order is considered. Another reason why WGC fails when applied to filtered signals is that the method becomes too conservative in detecting cross-frequency band causal relationships. This is evident from the cases of $Z^{\beta}_{\mathcal{Y}} \rightarrow Z^{\delta}_{\mathcal{X}}$ and $Z^{\theta}_{\mathcal{X}} \rightarrow Z^{\gamma}_{\mathcal{Y}}$ where WGC never detects these links across all simulated cases.

\begin{figure}
	\centerline{
		\includegraphics[width=0.85\textwidth]{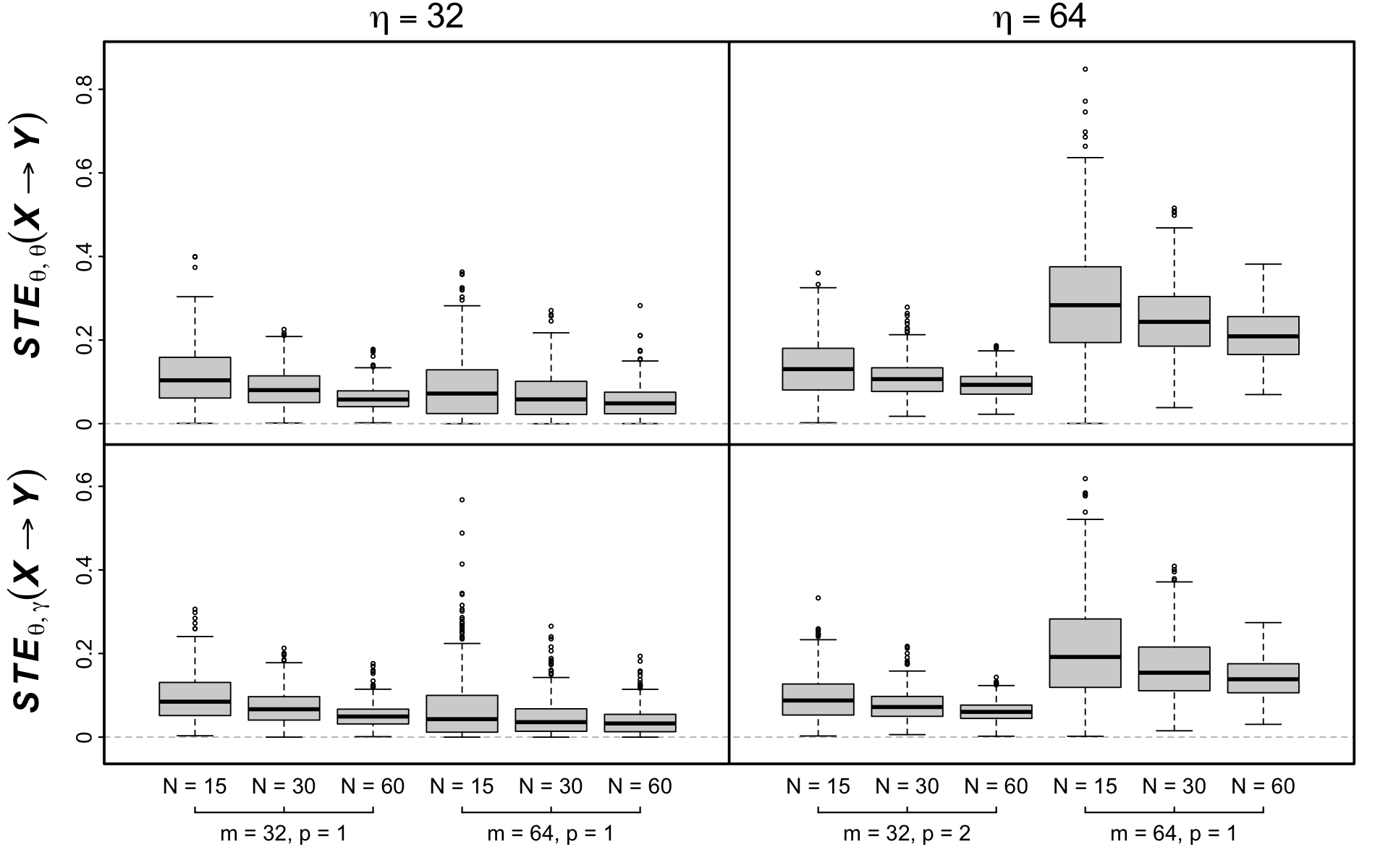}
	}
	\caption{Empirical distribution of the estimates for $\text{STE}_{\theta,\theta}(\mathcal{X} \rightarrow \mathcal{Y})$ and $\text{STE}_{\theta,\gamma}(\mathcal{X} \rightarrow \mathcal{Y})$ considering different lengths of observed signals ($N$) based from $500$ simulation replicates.}
	\label{fig:boxplot}
\end{figure}

Despite being defined on filtered band-specific signals, our STE approach is robust to such a phenomenon. In Tables~\ref{tab:sim_r1}~and~\ref{tab:sim_r2}, we highlight that our methodology is able to detect significant information flow with relatively high power while maintaining approximately correct sizes. That is, the proposed STE metric is able to capture causal links with controlled false detection rates. Similar to other procedures, our methodology also benefits from having more observations. Precisely, estimating STE over longer time series yields improvement in the power of the proposed test across all configurations of the tuning parameters $m$, $k$ and $\ell$. Furthermore, there is a visible reduction in the uncertainty of the STE estimates as $N$ increases, which roughly decreases proportional to the square root of the number of observations, as typically expected (see Figure~\ref{fig:boxplot}). This hints at the consistency of our estimation procedure and contributes to successfully detecting true causal links with high power. Such properties are desirable among statistical tools that make the STE framework appealing to be used in practice.

As anticipated, the correct choice of block size $m$ leads to the best performance of the STE-based causality framework. Specifically, when the simulated rate of information transfer is fast (i.e., $\eta = 32$), the proposed test attains its highest power when the chosen block size is $m = 32$ (see Table~\ref{tab:sim_r1}). In Table~\ref{tab:sim_r2}, we observe a similar behavior when the information transfer rate is slower (i.e., $\eta = 64$) and the STE metric is estimated with $m = 64$. In addition, the strength of causal relationship measured by the STE metric attains its maximum value with the appropriate choice of $m$ (see Figure~\ref{fig:boxplot}). However, the proposed framework achieves a sub-optimal performance when the specified $m$ does not match the rate of information transfer $\eta$. For instance, when $\eta = 32$, we observe less power when specifying $m = 64$ instead of the correct choice $m = 32$. Another effect of misspecified block size is that the estimated magnitude of information transfer is negatively biased. In Figure~\ref{fig:boxplot}, we see that the bias may be negligible when the information transfer with faster rate ($\eta = 32$) is estimated with larger block length ($m = 64$), i.e., the data is \textit{over-aggregated}, while the underestimation is more severe for the case when $\eta = 64$ but the STE metric is estimated with $m = 32$ (i.e., the data is \textit{under-aggregated}). One possible reason for this behavior is that under-aggregation of the oscillatory signals into block maxima of magnitudes introduces noise which results in the decrease in captured causal dependence. In contrast, over-aggregation translates to smaller effective sample sizes from where the STE metric is calculated, which then leads to an increase in uncertainty of the estimates. Considering this bias-variance trade-off, specifying larger values for $m$ is a more conservative choice to mitigate the bias, however, enough observations must be available to ensure that uncertainty of the estimates and the power of the test are within acceptable margins.

Along with the choice of $m$, the number of lag components $p$ (equivalent to $k$ and $\ell$ when $k=\ell$) included in the STE computation also affects the performance of our STE framework, especially the power of the proposed causality test. Since we simulate the modulating signals $A^{\Omega}_{\mathcal{X},t}$ and $A^{\Omega}_{\mathcal{Y},t}$, which carries the information being transferred, from an associated VAR($1$) process, we expect $p=1$ to be the optimal choice, given an appropriate $m$ has been specified. Expectedly, we find that the highest power among the configurations of STE is attained when $p = 1$ for the case of $\eta = 32$ given $m = 32$ and when considering $m = 64$ for the case of $\eta = 64$. In general, when there are more lag components included in the STE computation than the optimal number, the proposed test becomes more conservative resulting in reduced power. For both simulated rates of information transfer, we observe a decrease in power when considering higher lag orders for the STE metric. In the case of over-aggregation, i.e., choosing $m = 64$ when $\eta = 32$, the reduction in power is more prominent, especially for low sample size cases. When $\eta = 32$ and the optimal choice is $p = 1$, the information transfer occurs every quarter of a second. Estimating the STE measure with $m = 64$ and $p > 1$ means accounting for information transfer that occurs at least a whole second in the past, making the proposed test to be even more conservative. On the other hand, a strategic choice of $p$ can help improve the power of the test for cases of under-aggregation (i.e., when specifying $m < \eta$). For instance, in the simulated case where $\eta = 64$, the test based on an estimated STE with $m = 32$ and $p = 2$ performs almost at par with the optimal configuration of $m = 64$ with $p = 1$. As the information transfer occurs about every half a second when $\eta = 64$ with $p=1$ as the appropriate choice, considering $m = 32$ with $p = 2$ captures about the same temporal resolution for the causal influence, i.e., every two quarters of a second. Nonetheless, these results show the robustness of our method to the problems induced by linear filtering and provide evidence that the STE framework enables for causal inference on filtered band-specific signals.

Although an algorithm for the optimal selection of tuning parameters of the STE measure remains an open problem, we outline here some considerations when specifying values for $m$, $k$ and $\ell$ when using the proposed framework in practice. First, we need to decide on the temporal resolution at which the causality will be defined for the analysis, e.g., capturing information transfer that occurs within the past half second. This may be guided by context under study to achieve a desirable and interpretable resolution of causal influence. Then, find the balance between the block size and the number of lag components included in calculating the STE measure to match the specified resolution. If the goal of the analysis is to \textit{detect} presence of causal links between nodes in a network, we suggest lower values for the block size $m$ and adjust $k$ and $\ell$ accordingly. This ensures that the STE metric is estimated on larger effective sample sizes and improves the power of the causality test. Analysis of EEG recordings from trial-based experiments benefits from such configurations as the data available per trial are usually shorter in length. By contrast, if the goal of the analysis is to \textit{measure the magnitude} of information transfer, say to define dominant connections, we suggest choosing a more conservative value for $m$, $k$, and $\ell$, i.e., specifying a larger block size, and including less lag components in the STE computation. Estimates from this configuration are expected to have less bias, but require more observations to reduce the level of uncertainty and enhance the performance of the subsequent causality test. Causal connectivity investigations from continuously recorded EEG signals, similar to the data we analyze in this paper, may take advantage of these specifications for $m$, $k$ and $\ell$ since these kinds of data are often observed for longer temporal horizons. With this, we emphasize that our STE framework offers a novel yet easy-to-interpret approach for studying effective brain connectivity in the frequency domain.

\section{EEG Analysis: Brain Connectivity During Visual Task}\label{chap:analysis}

\subsection{Scientific Problem and Summary of the Analysis}

The goal of this paper is to characterize the effective brain connectivity, during cognition, of children diagnosed with attention deficit hyperactivity disorder (ADHD), in comparison to children without any registered psychiatric disorder (healthy controls).  Specifically, we want to identify causal relationships between brain regions of subjects with ADHD that significantly differ from a healthy subject's connectivity. Another objective is to establish the connection of such relationships to specific frequency bands that have well-known associated cognitive functions. For these purposes, existing methods are inadequate because they are based on too simplistic or too restrictive assumptions. Moreover, these approaches construct the causal structure in the ``frequency-specific'' paradigm which, most of the time, is difficult to interpret and has potentially inflated false rejection rates. Thus, we here apply our new spectral causal STE framework to address these scientific questions. 

\begin{figure}
	\centerline{
		\includegraphics[width=\textwidth]{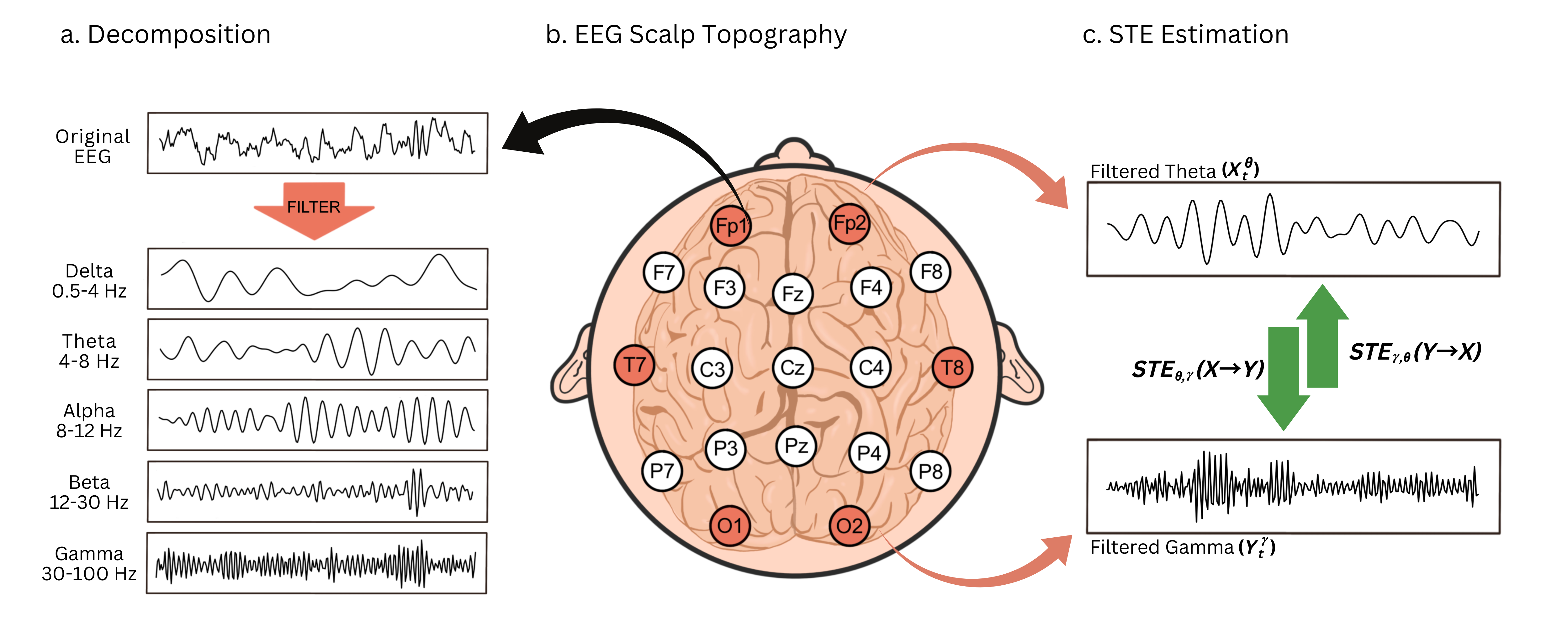}
	}
	\caption{Building blocks of STE. (\textbf{a}) Application of bandpass filter on original EEG series to obtain frequency band-specific filtered series. (\textbf{b}) Standard 10--20 EEG scalp topography where the highlighted channels are included in the analysis. (\textbf{c}) Illustration of STE estimation between two filtered series.}
	\label{fig:eegSTE}
\end{figure}

We analyze EEG recordings of two groups of children (the ADHD group and healthy control group) during performance of a visual task. The data, collected by \cite{rzfh-zn36-20}, contains samples of EEGs at 128Hz, from 19 channels of 51 children with ADHD and 53 healthy subjects. Average recordings from channels A1 and A2 (located at the two earlobes) were used as electrode references. Pre-processing was conducted via the PREP pipeline of \cite{bigdely2015prep} to increase the quality of the recordings and remove unwanted artifacts due to electrical interference and muscle movements including eye blinking and ear twitching. The visual-cognitive experiment was to count the number of characters in a flashed image (recall Figure~\ref{fig:eeg_visual}). Hence, to ``represent'' brain regions that are highly likely to be engaged during the cognitive task, we selected six EEG channels, namely, Fp1 (left pre-frontal), Fp2 (right pre-frontal), T7 (left temporal), T8 (right temporal), O1 (left occipital) and O2 (right occipital) (see Figure~\ref{fig:eegSTE}b). Given that the frontal region is linked with concentration, focus and problem solving, the temporal region with speech and memory, and the occipital region with visual processing \citep{bjorge2017identification}, the interest now is to identify (in both the ADHD and control groups) which cross-channel information transfer is significant and at which frequency bands it occurs.

During the experiment, the succeeding images are flashed right after the subject answers how many characters are in the shown picture. This ensures that the EEG recordings reflect the continuous thinking process associated to the visual task without any interruptions. Here, we analyze recordings after the first ten seconds of the experiment to minimize inclusion of signals that are unstable and may be unrelated to visual-attention cognition. However, some subjects take a longer time to complete the visual experiment, and this thus results in varying lengths of available EEG recordings, ranging from 45 to 275 seconds. To obtain the filtered frequency band-specific series per channel, fourth order Butterworth band-pass filters are applied to the pre-processed EEG series (see Figure~\ref{fig:eegSTE}a). Then, we calculate the STE for each possible pair of the selected channels and for each pair of possible frequency bands (see Figure~\ref{fig:eegSTE}c). That is, for all $i,j = 1, \ldots, 6$, $i \neq j$ and $\Omega_1, \Omega_2 \in \{\delta,\theta,\alpha,\beta,\gamma\}$, we estimate $STE_{\Omega_1,\Omega_2}(C_i \rightarrow C_j;k,\ell)$ individually for each subject, where $C_1$, $C_2$, $C_3$, $C_4$, $C_5$ and $C_6$ denote the recordings from channels Fp1, Fp2, O1, O2, T7, and T8, respectively. 

In the subsequent STE estimation, we consider the block size $m = 64$, which is equivalent to aggregating the magnitude of the filtered signals every half second, and let the number of lag components of the STE metric to be $k = \ell = 2$, suggesting the information transfer to occur within a one-second period. As neuronal transmission of information between brain regions may last several hundreds of milliseconds, e.g., word processing can take up to 800 milliseconds \citep{schack2003cerebral} while simple visual categorization takes around 450 milliseconds \citep{kirchner2006ultra}, we opt for these conservative values for $m$, $k$, and $\ell$ to account for the neurophysical nature of the brain. In the Supplementary Material, we report results of our sensitivity analysis where we set instead $m = 32$ with $k = \ell = 3$ or $m = 16$ with $k = \ell = 4$, which generally points to similar interpretations as the findings in this paper. In addition, we address issues of non-stationarity in the STE estimation by employing the strategy of fitting marginal distributions via the ECDF over locally stationary time series segments of length 15 seconds, as discussed in Section~\ref{subchap:margins}. Then, tests for significance of the calculated STE values are conducted based on $5000$ resamples. Since there are $\binom{6}{2} = 15$ channel pairs, $5 \times 5 = 25$ frequency band pairs, and $2$ causal directions, a total of $750$ individual significance tests for each of the $104$ subjects are implemented and their corresponding BH-adjusted p-values are used to describe the brain connectivity of the two groups.

For ease of interpretation, we aggregate the causal relationships quantified by our STE framework into two broader types of connectivity, namely, directional links from \textit{low-frequency sender signals} and from \textit{high-frequency sender signals}. Specifically, we let
\begin{align*}
    STE_L(C_i \rightarrow C_j) & = \max_{\Omega_1 \in \{\delta,\theta,\alpha\}} \big\{ STE_{\Omega_1,\Omega_2}(C_i \rightarrow C_j)\big\}, \\
    \text{and}~~~ STE_H(C_i \rightarrow C_j) & = \max_{\Omega_1 \in \{\beta,\gamma\}} \big\{ STE_{\Omega_1,\Omega_2}(C_i \rightarrow C_j)\big\}.
\end{align*}
Connections involved with $STE_L(C_i \rightarrow C_j)$ are information pathways from channel $C_i$ to $C_j$ where low frequencies (including the delta, theta and alpha bands) of $C_i$ ``cause'' the oscillations of $C_j$. As the cognitive functions required to sustaining attention and filtering out irrelevant information during performance of a mental task are related to these low frequencies \citep{fernandez1995eeg,harmony1996eeg,klimesch1998induced,clayton2015roles,behzadnia2017eeg,helfrich2018neural,van2019functional,zhang2022default}, we interpret such connections, originating from low-frequency sender signals, as crucial drivers for keeping focus and maintaining attention. On the other hand, $STE_H(C_i \rightarrow C_j)$ comprises of directional links from channel $C_i$ to $C_j$ where high frequency oscillations (involving the beta and gamma bands) of $C_i$ transfers information to the oscillatory components of $C_j$. Because beta and gamma oscillations relate to cognitive processing and informational integration \citep{miltner1999coherence,muller2000modulation,jensen2007human,she2012eeg,gola2013eeg,strube2021alpha}, we interpret these connections originating from the high-frequency sender signals, as potential markers that are actively engaged during formulation of thought processes, e.g., the conscious mental activity to carry out the visual task of counting characters in a flashed image.

For each subject, we then derive the relevant effective brain connectivity networks based on \textit{significant} and \textit{dominant} $STE_L$ and $STE_H$ values. To be more precise, we consider $STE_L(C_i \rightarrow C_j)$ to be significant if, for at least one pair $(\Omega_1,\Omega_2)'$ where $\Omega_1 \in \{\delta,\theta,\alpha\}$, $STE_{\Omega_1,\Omega_2}(C_i \rightarrow C_j)$ is significant, i.e., if at least one of the corresponding BH-adjusted p-values of $STE_{\Omega_1,\Omega_2}(C_i \rightarrow C_j)$ is less than the $0.10$ level of significance. Across all subjects, the smallest and largest non-zero BH-adjusted p-values, which are considered significant, are $0.0103$ and $0.0997$, respectively (with unadjusted values of $0.0002$ and $0.0104$, respectively). However, since the STE metric does not differentiate between direct connections (e.g., Fp1~$\rightarrow$~O1) and indirect connections (e.g., Fp1~$\rightarrow$~T7~$\rightarrow$~O1), generating the brain network based solely on significant STE values may involve indirect connections, and thus, we further consider connections with dominant magnitudes. Specifically, we define the directional link $C_i \rightarrow C_j$, originating from low-frequency sender signal, to be dominant if $STE_L(C_i \rightarrow C_j) > \underset{i \neq j}{\mathrm{median}} \big\{STE_L(C_i \rightarrow C_j)\big\}$, i.e., if the magnitude of the information flow from $C_i$ to $C_j$ belongs to the top 50-$th$ percentile of causal links across all $i$ and $j$. By considering dominant directional links, assuming direct connections have larger magnitudes compared to indirect connections, we limit the inclusion of the latter in the derived brain network for a specific subject. Similar concepts of significance and dominance are defined for the links involving high-frequency sender signals based on $STE_H$.

Finally, we summarize the brain connectivity networks related to maintaining attention and information processing for the ADHD group and the control group by considering all directional links detected by our STE framework, that are both significant and dominant in at least 50\% of the subjects in each group. These connections are reported in Table~\ref{tab:derived} and the corresponding networks are displayed graphically in Figures~\ref{fig:EEGSumm_low}~and~\ref{fig:EEGSumm_high}. Individual connectivity networks for some subjects, are also discussed with respect to the aggregated group networks in the Supplementary Material.

\begin{table}
\caption{Summary of significant and dominant information transfer originating from low-frequency and high-frequency sender signals detected by the proposed STE framework from at least $50\%$ of subjects in the ADHD group and in the control group.}
\label{tab:derived}
\begin{center}
\begin{tabular}{lcccc||lcccc}
\multirow{2}{*}{\textbf{Causal Link}} & \multicolumn{2}{c}{\textbf{ADHD}} & \multicolumn{2}{c}{\textbf{Control}} &  \multirow{2}{*}{\textbf{Causal Link}} & \multicolumn{2}{c}{\textbf{ADHD}} & \multicolumn{2}{c}{\textbf{Control}}\\
 & \textbf{Low} & \textbf{High} & \textbf{Low} & \textbf{High} & & \textbf{Low} & \textbf{High} & \textbf{Low} & \textbf{High}\\
\hline
\hline              
$Fp1 \rightarrow Fp2$ &                & $\usym{1F5F8}$ & $\usym{1F5F8}$ &                & $Fp2 \rightarrow Fp1$ & $\usym{1F5F8}$ & $\usym{1F5F8}$ & $\usym{1F5F8}$ &               \\

$Fp1 \rightarrow O1$ & $\usym{1F5F8}$ &                & $\usym{1F5F8}$ &                & $Fp2 \rightarrow O1$ & $\usym{1F5F8}$ &                & $\usym{1F5F8}$ &                \\

$Fp1 \rightarrow O2$ &                & $\usym{1F5F8}$ & $\usym{1F5F8}$ &                & $Fp2 \rightarrow O2$ &                &                & $\usym{1F5F8}$ &               \\

$Fp1 \rightarrow T7$ & $\usym{1F5F8}$ &                &                &                & $Fp2 \rightarrow T7$ & $\usym{1F5F8}$ &                & $\usym{1F5F8}$ & $\usym{1F5F8}$\\

$Fp1 \rightarrow T8$ & $\usym{1F5F8}$ &                & $\usym{1F5F8}$ & $\usym{1F5F8}$ & $Fp2 \rightarrow T8$ &                &                &                &               \\

\hline

$O1 \rightarrow Fp1$ &                & $\usym{1F5F8}$ &                &                & $O2 \rightarrow Fp1$ &                &                &                &               \\

$O1 \rightarrow Fp2$ &                & $\usym{1F5F8}$ &                &                & $O2 \rightarrow Fp2$ &                &                &                &               \\

$O1 \rightarrow O2$ &                & $\usym{1F5F8}$ &                &                & $O2 \rightarrow O1$ &                &                &                &               \\

$O1 \rightarrow T7$ &                & $\usym{1F5F8}$ &                & $\usym{1F5F8}$ & $O2 \rightarrow T7$ &                &                &                & $\usym{1F5F8}$\\

$O1 \rightarrow T8$ &                & $\usym{1F5F8}$ &                & $\usym{1F5F8}$ & $O2 \rightarrow T8$ &                & $\usym{1F5F8}$ &                &               \\

\hline

$T7 \rightarrow Fp1$ &                & $\usym{1F5F8}$ &                &                & $T8 \rightarrow Fp1$ &                &                & $\usym{1F5F8}$ &               \\

$T7 \rightarrow Fp2$ &                &                &                &                & $T8 \rightarrow Fp2$ &                &                & $\usym{1F5F8}$ & $\usym{1F5F8}$\\

$T7 \rightarrow O1$ & $\usym{1F5F8}$ &                & $\usym{1F5F8}$ & $\usym{1F5F8}$ & $T8 \rightarrow O1$ & $\usym{1F5F8}$ &                & $\usym{1F5F8}$ &               \\

$T7 \rightarrow O2$ &                &                & $\usym{1F5F8}$ &                & $T8 \rightarrow O2$ & $\usym{1F5F8}$ &                & $\usym{1F5F8}$ &               \\

$T7 \rightarrow T8$ & $\usym{1F5F8}$ & $\usym{1F5F8}$ & $\usym{1F5F8}$ & $\usym{1F5F8}$ & $T8 \rightarrow T7$ & $\usym{1F5F8}$ &                & $\usym{1F5F8}$ & $\usym{1F5F8}$\\

\hline
\hline
\end{tabular}
\end{center}
\end{table}

\subsection{Results: Effective Causal Network related to Maintaining Attention}

\begin{figure}
	\centerline{
		\includegraphics[width=0.55\textwidth]{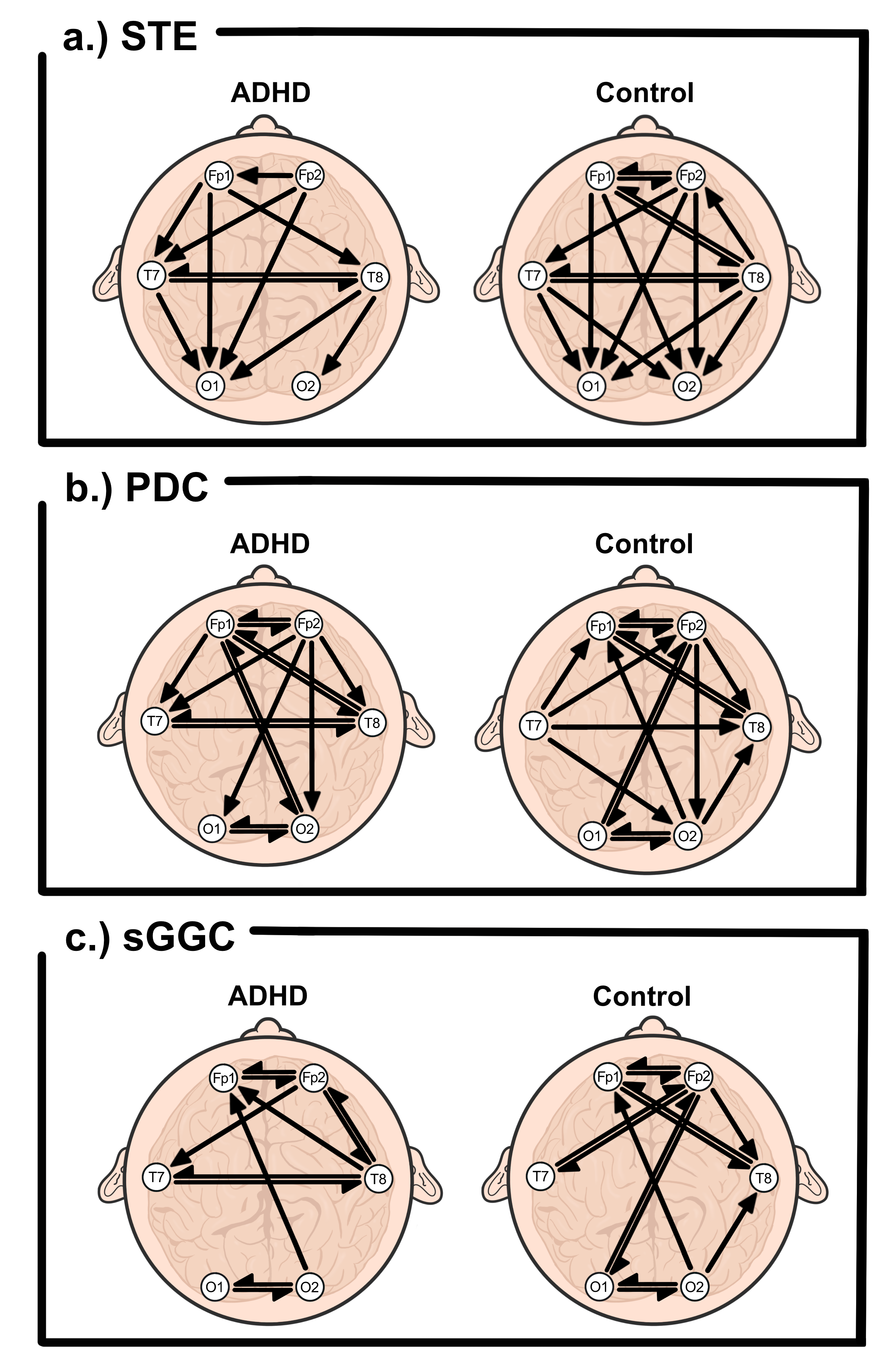}
	}
	\caption{Derived aggregated effective connectivity networks involving low-frequency sender signals (related to maintaining attention) for the ADHD and control groups based on three causal frameworks: the proposed STE (top), PDC (middle), and sGGC (bottom).}
	\label{fig:EEGSumm_low}
\end{figure}

In Figure~\ref{fig:EEGSumm_low}, we illustrate the derived brain networks for the ADHD group and the healthy control group, involving significant and dominant connections that originates from low-frequency sender signals. As low frequency signals are associated with the ability to sustain attention, one of our prior speculations is the presence of information transfer from the pre-frontal to the occipital channels. That is, we suspect that the ability to concentrate and keep focus during the visual task requires some flow of information from the problem-solving channels (Fp1 and Fp2) to the visual-processing channels (O1 and O2). Such connections may occur directly (e.g., Fp1~$\rightarrow$~O1) or indirectly through another channel (e.g., Fp1~$\rightarrow$~T8~$\rightarrow$~O1). These may be equivalent to the cognitive function of concentrated (visual) observation which is needed when counting characters in an image.

Our STE framework reveals that, for the healthy control groups, the information pathways from the pre-frontal to the occipital channels are highly dense. For instance, all possible direct pathways of information flow from channels Fp1 and Fp2 to the channels O1 and O2 are deemed significant and dominant. In addition, numerous indirect links, which passes through the temporal channels, are observed among healthy subjects, e.g., Fp1~$\rightarrow$~T8~$\rightarrow$~O1, Fp2~$\rightarrow$~T7~$\rightarrow$~O2, and Fp1~$\rightarrow$~T8~$\rightarrow$~T7~$\rightarrow$~O2. We also observe a two-way feedback mechanism between the two problem-solving pre-frontal channels, namely, Fp1~$\rightarrow$~Fp2 and Fp2~$\rightarrow$~Fp1, and connections from the temporal to the pre-frontal channels, i.e., T8~$\rightarrow$~Fp1 and T8~$\rightarrow$~Fp2. With the exemption of the directional links from channel T8, the derived brain network related to maintaining attention for the healthy control group is highly symmetric with respect to the brain hemispheres. Along with the abundance of information pathways from the pre-frontal to the occipital channels, such symmetry in connectivity structure may be the essential drivers that help the healthy subjects in keeping a longer and more stable attention capacity.

In contrast, although several similar pathways are detected among the ADHD subjects, we observe significantly less directional links from the pre-frontal to the occipital channels in the derived network for the ADHD group. Another difference is that the problem-solving channels Fp1 and Fp2 do not have a two-way feedback mechanism, i.e., Fp2~$\rightarrow$~Fp1 but not Fp1~$\rightarrow$~Fp2. Moreover, with respect to the left and right brain hemispheres, the effective network of the ADHD group appear to be asymmetric, which we interpret as instabilities in the connectivity structure. We speculate that the missing information pathways from the pre-frontal to the occipital channels, as well as the asymmetry in their network, may be linked to ADHD subjects having known shorter attention span.

\subsection{Results: Effective Causal Network related to Information Processing}

\begin{figure}
	\centerline{
		\includegraphics[width=0.55\textwidth]{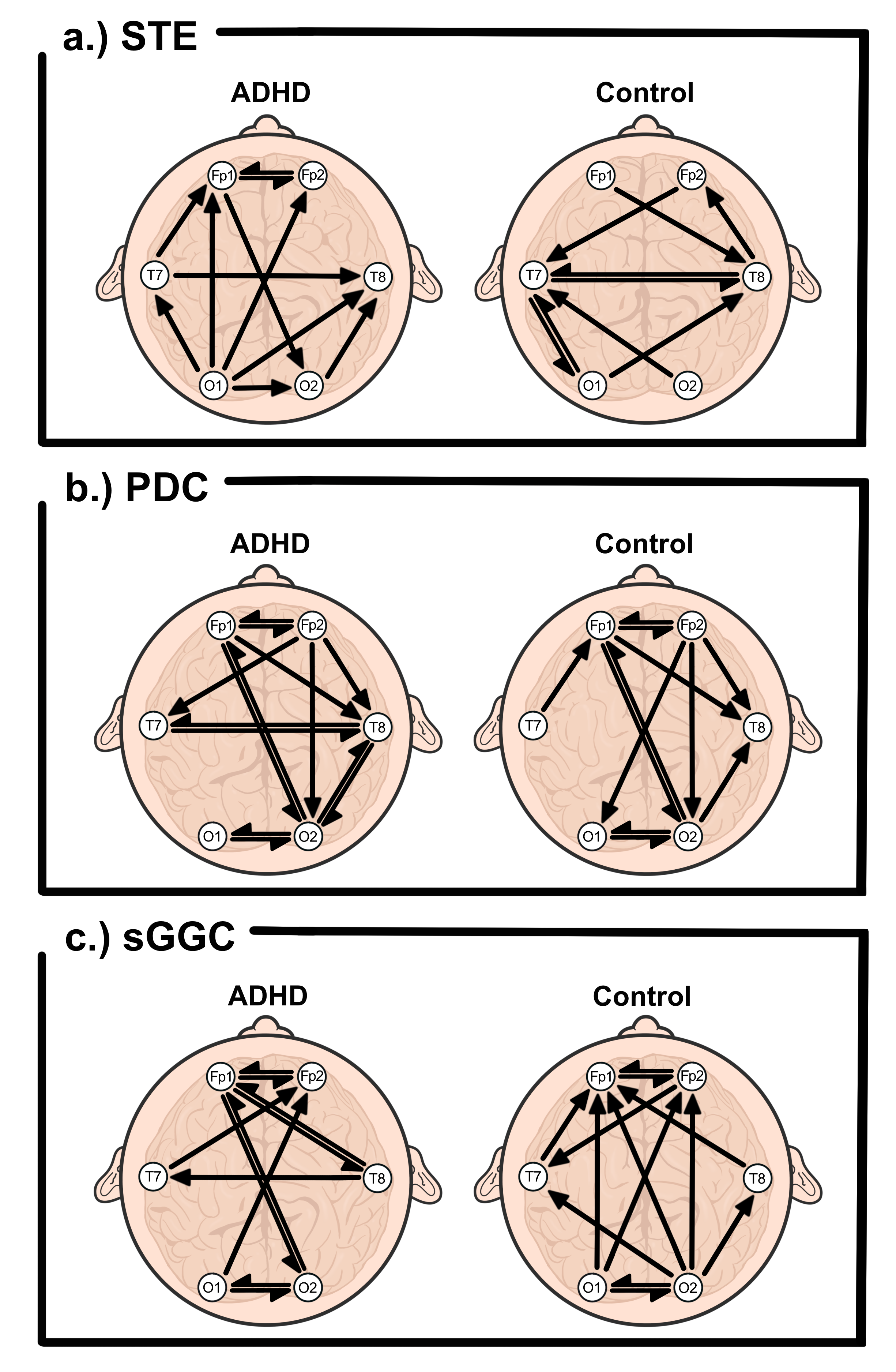}
	}
	\caption{Derived aggregated effective connectivity networks involving high-frequency sender signals (related to information processing) for the ADHD and control groups based on three causal frameworks: the proposed STE (top), PDC (middle), and sGGC (bottom).}
	\label{fig:EEGSumm_high}
\end{figure}

Figure~\ref{fig:EEGSumm_high} illustrates the derived connectivity networks, which involve significant and dominant directional links originating from high-frequency sender signals, for the ADHD group and the healthy control group. Associated with complex information processing, we interpret connections from these networks as the information pathways necessary to count visual images. We hypothesize that the said thinking process requires some form of interaction between the occipital channels, at which visual inputs are received, and the pre-frontal channels where mental calculations may take place. 

Results from our STE approach suggest the active utilization of temporal channels among the healthy controls. Specifically, we observe significant and dominant information flow from the pre-frontal to the temporal channels, i.e., Fp1~$\rightarrow$~T8 and Fp2~$\rightarrow$~T7, and from the occipital to the temporal channels, i.e., O1~$\rightarrow$~T8 and O2~$\rightarrow$~T7. Together with the two-way feedback between memory channels T7 and T8, we speculate that the flow of information converging towards the temporal channels resembles the phenomenon of image recall. That is, whenever similar images are shown, it becomes easier for healthy subjects to perform the visual experiment. This is evident from the faster completion time of the healthy control group with an average and standard deviation of 121.53 seconds and 29.36 seconds, respectively, as compared to the ADHD group, with an average of 149.35 seconds and standard deviation of 55.22 seconds.

Another characteristic observed in the derived brain network for the control group is the presence of multiple continuous loops of information pathways. For instance, the pre-frontal and the temporal channels exhibit such continuous loop through Fp2~$\rightarrow$~T7~$\rightarrow$~T8~$\rightarrow$~Fp2, as well as the occipital and the temporal channels through O1~$\rightarrow$~T8~$\rightarrow$~T7~$\rightarrow$~O1. In addition, the detected pathway O1~$\rightarrow$~T8~$\rightarrow$~Fp2~$\rightarrow$~T7~$\rightarrow$~O1 suggests a continuous form of information transfer between the pre-frontal and the occipital channel which passes through the temporal channels. This provides further evidence in support of our conjecture that the healthy subjects utilize the temporal channels during cognition of counting characters. Assuming that the brain is an efficient entity, the connectivity patterns visualized among healthy controls resemble a simpler but more systematic brain wiring related to carrying out visual tasks.

By contrast, a different connectivity behavior related to information processing is observed among subjects with ADHD. In comparison to healthy controls, we observe a two-way feedback between channels Fp1 and Fp2 in the derived network for the ADHD group. As the pre-frontal channels are associated with concentration, information transfer involving high-frequency sender signals in the said brain region may be indicative of the active and conscious efforts among ADHD subjects to maintain focus, compensating for their shorter attention span. Moreover, the flow of information in the derived brain network for the ADHD group is evidently less systematic. For example, channels with similar associated cognitive functions do not necessarily have the same role. In the temporal areas, channel T7 serves as the intermediate node in the indirect information pathway from the occipital to the pre-frontal channel (i.e., O1~$\rightarrow$~T7~$\rightarrow$~Fp1), while channel T8 only receives information from other channels (i.e., O1~$\rightarrow$~T8, O2~$\rightarrow$~T8, and T7~$\rightarrow$~T8). Between the two occipital channels, information flows from channel O1 to all other five channels (i.e., O1~$\rightarrow$~O2, O1~$\rightarrow$~Fp1, O1~$\rightarrow$~Fp2, O1~$\rightarrow$~T7, and O1~$\rightarrow$~T8), in contrast to channel O2 receiving information from channel Fp1. We interpret these channel-role differences as instabilities in information processing which may be a by-product of inattentiveness innate to ADHD patients.

\subsection{Results: Complementing Frequency-specific Causal Frameworks}

To our knowledge, the proposed STE framework is the first to investigate causal interactions between nodes of a brain network in the ``frequency band''-specific paradigm. This enables for capturing relatively easy-to-interpret effective connectivity patterns that can be associated to known cognitive functions. Moreover, our methodology is not limited only to oscillations coming from the same frequency band, and thus, enables for detecting cross-frequency interactions. However, due to aggregating signals into series of block maxima of magnitudes, the temporal resolution of causal influence measured by STE is lower than the original time scale of the observed data. Therefore, our STE approach complements (rather than competes against) other existing ``frequency''-specific causal frameworks, and hence, it helps provide a better understanding of the underlying causal dynamics in the brain.

For completeness, we now compare our findings to two commonly-used frequency-specific methodologies for effective brain connectivity \citep{cekic2018time}. Specifically, we analyze the same EEG data by implementing the partial directed coherence (PDC) of \cite{baccala2001partial} and the spectral conditional Geweke-Granger causality (sGGC) measure of \cite{geweke1984measures} to detect \textit{dominant} connections and derive congruent effective connectivity networks for the ADHD and control groups. Defined from the Fourier transform of the coefficients of a stationary VAR($p$) process, the PDC measure reflects the direction and magnitude of information flow, at a specific individual frequency, from one channel to another, relative to all other information transfer coming from the same source. On the other hand, the sGGC measure is the frequency decomposition of the Granger causality measure derived from corresponding restricted and unrestricted VAR($p$) processes. We refer to the original works of \cite{baccala2001partial} and \cite{geweke1984measures} for the formulation and estimation of their respective metrics. Because both PDC and sGGC measures are obtained from a fitted VAR model, the causal structure they capture occurs at the same temporal scale as the original data, e.g., in milliseconds depending on the sampling rate at which the signals were recorded. Thus, the two approaches account for linear intra-frequency interactions occurring at "almost instantaneous" rates, in contrast to our STE measure that accounts for non-linear cross-frequency information transfer but at slower (thus, neurologically more meaningful) rates.

In Figures~\ref{fig:EEGSumm_low}~and~\ref{fig:EEGSumm_high}, we illustrate the derived effective networks related to low and high frequencies based on dominant PDC and sGGC values (see the Supplementary Material for details on the implementation). For the ADHD and control groups, PDC and sGGC suggest a two-way feedback between the pre-frontal channels (i.e., Fp1~$\leftrightarrow$~Fp2), as well as between the occipital channels (i.e., O1~$\leftrightarrow$~O2). These connections are detected at low and high frequencies which support the contribution of the problem-solving (pre-frontal) channels and the visual processing (occipital) channels in performing the experiment of counting images.

However, it is difficult to understand the neurological dysfunctions associated with ADHD solely based on the PDC and sGGC frameworks. This is because the two approaches identify minimal differences in the effective networks between the two groups of subjects. For instance, based on PDC, the major role of channel T7 is to receive information for the ADHD group while it sends information to other channels for the control group. Moreover, sGGC detects a two-way low-frequency interaction between the two temporal channels in the ADHD group but not among the healthy controls. Since the temporal channels are linked with the cognition of speech and memory, it is unclear how these discrepancies may differentiate the capacity to maintain attention of healthy subjects from subjects with ADHD. Nevertheless, sGGC is able to identify a missing two-way link between channels Fp2 and O1 from the ADHD group, which may be related to having shorter attention span.

At high frequencies, we observe some disagreements between the effective networks related to information processing based on PDC and sGGC. For the former, the flow of information generally occurs from the pre-frontal to the occipital channels, which is the opposite for the latter. Moreover, PDC identifies presence of a two-way information transfer between the temporal channels only in the ADHD group and a causal link from channel Fp2 to O1 only among the healthy control group. By contrast, additional directional links from the occipital to the pre-frontal channels (i.e., O1~$\rightarrow$~Fp1 and O2~$\rightarrow$~Fp2), and information flow towards the temporal channels (i.e., Fp2~$\rightarrow$~T7, O2~$\rightarrow$~T7 and O2~$\rightarrow$~T8) differentiate the control group from the ADHD group, as suggested by the sGGC framework. Nonetheless, combining these results with the novel findings based on our STE framework is useful in fully characterizing the spectral causal structure in the brain during the visual experiment and helps better understand the neurological dysfunctions associated with ADHD.

\section{Conclusion}\label{chap:conclusion}

During the performance of a specific cognitive task, interactions between different brain regions are characterized in the form of information transfer. When an individual is diagnosed with neuro-developmental disorder (e.g., ADHD), information flow between some brain regions may be altered (e.g., completely disupted or weakened) compared to a healthy individual, thus providing explainable brain connectivity pathways for the patient population with cognitive dysfunctions. In EEG analysis, this phenomenon can be visualized through the strength of cross-channel information flow. Hence, it is essential to have a quantitative metric that adequately captures both the magnitude and direction of this information transfer between the nodes in a brain network.

Transfer entropy (TE), an information-theoretic measure that offers a general framework for quantifying the flow of information between two channels. As several brain functions are attributed to specific groups of frequency oscillations (frequency bands), application of TE in the frequency domain is desirable. Thus, we have developed a new causal metric called spectral transfer entropy (STE) by calculating TE on non-overlapping block maxima series computed from the magnitude of filtered ``frequency band-specific'' signals.

The advantage of our proposed STE metric is that it provides the magnitude of information transfer from a specific oscillatory component of a series to another oscillatory component of another series. It is useful to clinicians since its interpretation is directly linked to standard frequency bands, unlike other works on TE in the frequency domain (which are ``frequency-specific'' for individual frequencies, rather than ``frequency band-specific''). Furthermore, our proposed methodology easily allows for adjustment for multiple comparisons which are often not addressed in its frequency-specific counterparts. With this, the risk of detecting spurious information transfer is no longer an issue since controlling for the false positive rate is straightforward.

In addition, we propose a novel estimation method for TE/STE that exploits a specific vine copula structure leading to robust and computationally efficient inference. Given that TE/STE can be expressed through the conditional mutual information (CMI), calculations for TE/STE are simplified by considering an appropriate D-vine structure for the full joint copula density. One advantage of this approach is that it leads to simpler and more efficient way of estimating TE/STE based on a single fitted model that can be used for both causal directions. Another advantage of our approach is that it is able to obtain an exact ``zero'' TE/STE, which removes the necessity of adjusting the estimates for bias. Moreover, measuring uncertainty of the estimates based on a standard resampling scheme is fast because generating new observations from an estimated vine copula is straightforward. Numerical experiments illustrate that our inference approach attains correct sizes and high power, providing evidence on the robustness of STE to the issues imposed by filtering.

Our novel STE framework provides both interesting and novel findings in the analysis of EEG data linked to a visual task. In relation to maintaining attention, missing information pathways from the pre-frontal to the occipital channels, linked with low frequency oscillations, differentiate the ADHD group from the control group. This may be related to ADHD subjects having shorter attention span. Our proposed STE method has also uncovered a novel finding on brain networks associated with high-frequency oscillations. Compared to the ADHD group, the derived brain network for the control group reveals a more systematic connectivity pattern that utilizes the temporal channels in cognitive information processing, hence explaining the efficiency of healthy subjects in performing the experiment.

Even though the STE measure is developed for oscillatory processes, it could also be used for any other time series (e.g., financial data), although it may then be more appropriate to call it a ``tail'' transfer entropy (TTE) metric, since it measures the amount of information transferred from the tail of the distribution (through block maxima) of one time series to another series' tail. For applications that require looking at block minima or a combination of block maxima and block minima, e.g., in modeling extremal causal relationships in stock prices, this can be straightforwardly done in the proposed framework, as well.

Lastly, since the STE measure uses the maximum magnitude over time blocks to capture information transfer between oscillatory processes, a natural extension is to consider using high threshold exceedances instead. That is, utilize the data from all time points where the value of the series exceeds some high threshold. However, an issue with this approach is to define the temporal scale of information transfer because exceedances do not happen on equally-spaced time intervals which leads to difficulty in  establishing ``how much of the past causes the present''. Nonetheless, combining extreme value theory with modeling causal relationships will be a new and novel approach, which may address existing issues of standard methods in understanding brain connectivity.






\begin{acks}[Acknowledgments]
The authors thank Sarah Bernadette Aracid for the elaborate artworks (Figures~\ref{fig:eeg_visual}, \ref{fig:vinecopula}, \ref{fig:resampTE}, \ref{fig:eegSTE}, \ref{fig:EEGSumm_low}, and \ref{fig:EEGSumm_high}).
\end{acks}



\begin{supplement}
\stitle{Supplementary Material for ``Measuring Information Transfer Between Nodes in a Brain Network through Spectral Transfer Entropy''}
\sdescription{Includes the derivations for the simplification of TE via the vine copula representation when $k = \ell = 2$ and $k = \ell = 3$, the specifications of the models/metrics used for the numerical experiments and complementary data analysis, some outputs from subject-specific analysis, and the results of the sensitivity analysis for the choice of signal filter and tuning parameters of the STE measure for the analysis of EEG data.}
\end{supplement}


\bibliographystyle{imsart-nameyear} 
\bibliography{bibliography}       

\end{document}